\pdfoutput=1

\documentclass[aps,twocolumn,showpacs,amsmath,amssymb,nofootinbib,superscriptaddress,showkeys,prd,floatfix]{revtex4-1}

\usepackage{epsfig}
 \usepackage[section]{placeins}
 
\newcommand{\br}[1]{\langle #1\rangle}

\begin{document} 

\title{Forward-backward multiplicity fluctuations in ultra-relativistic nuclear collisions \\ with wounded quarks and fluctuating strings}

\author{Martin Rohrmoser}
\email{mrohrmoser@ujk.edu.pl} 
\affiliation{Institute of Physics, Jan Kochanowski University, 25-406 Kielce, Poland}

\author{Wojciech Broniowski}
\email{Wojciech.Broniowski@ifj.edu.pl}
\affiliation{Institute of Physics, Jan Kochanowski University, 25-406 Kielce, Poland}
\affiliation{H. Niewodnicza\'nski Institute of Nuclear Physics PAN, 31-342 Cracow, Poland}

\date{21 September 2018}  

\begin{abstract}
We analyze a generic model where wounded quarks are amended with strings in which both end-point positions fluctuate 
in spatial rapidity. With the assumption that the strings emit particles independently of one another and with a uniform
distribution in rapidity, we are able to analyze the model semi-analytically, which allows for its detailed understanding.
Using as a constraint the one-body string emission functions obtained from the experimental data for collisions at $\sqrt{s_{NN}}=200$~GeV, 
we explore  the two-body correlations for various scenarios of string fluctuations. 
We find that the popular measures used to quantify the longitudinal fluctuations ($a_{nm}$ coefficients) are limited with upper and lower bounds.
These measures can be significantly larger in the model where both end-point are allowed to fluctuate, compared to the model with 
single end-point fluctuations.
\end{abstract}


\maketitle

\section{Introduction \label{sec:intro}}

The purpose of this paper is to present a detailed semi-analytic analysis of models of ultra-relativistic 
nuclear collisions where the early production of particles occurs from strings.
The strings are associated with wounded quarks, and both of their end-point positions fluctuate 
in spatial rapidity. The model generalizes the analysis of~\cite{Broniowski:2015oif} where only one-end fluctuations were 
considered.
The main assumptions are that the strings emit particles independently of one another and 
that the production from a string is uniform between its end-points.
We obtain the one-body string emission function from a fit to the experimental data at $\sqrt{s_{NN}}=200$~GeV, 
and use it to constrain the freedom in the distribution of the end-point positions. We  then
explore in detail the two-body correlations in various scenarios for the fluctuating end-points. 
The derived analytic formulas allow for a full understanding of this simple model. In particular,
we show that standard measures applied in analyses of the longitudinal fluctuations, 
such as the Legendre $a_{nm}$ coefficients, fall between certain bounds. This 
explains why a priori different models may provide quite similar results for these measures of the longitudinal correlations. 
We find that the $a_{nm}$ coefficients can be significantly larger (by a factor of $\sim 3$) when one allows for two end-point to fluctuate, 
compared to the case of single end-point fluctuations of~\cite{Broniowski:2015oif}. This observation is 
relevant for phenomenological studies. 
Since the model, despite its simplifications, is generic, sharing features with more complicated 
string implementations, our findings shed light on correlations from other string models in application to 
ultra-relativistic heavy-ion collisions.

The basic phenomenon explored in this paper and illustrated with definite calculations can be understood in very simple terms. 
Consider a string with left and right end-points and an acceptance window in pseudorapidity. If the left end-point 
were always left of the acceptance window, 
and the right end-point to the right (they may fluctuate or not, but cannot enter the window), then the string seen in the 
window is always the same, hence no fluctuations occur. If, 
however, an endpoint via fluctuation enters the acceptance 
window, then fluctuations occur, as its observed fragment may 
be shorter or longer. The fluctuation effect is larger when both end-points fluctuate 
into the acceptance window, which is the case explored in detail below.

The concept of wounded sources formed in the initial stages of ultra-relativistic heavy-ion collisions has proven to be phenomenologically successful in 
reproducing multiplicity distributions from soft particle production. The idea (see~\cite{Bialas:2008zza} for 
a discussion of the foundations), adopts the Glauber model~\cite{Glauber:1959aa} in its variant suitable for inelastic collisions~\cite{Czyz:1969jg}. 
Whereas the wounded nucleon scaling~\cite{Bialas:1976ed}, when applied to the highest BNL Relativistic Heavy-Ion Collider (RHIC) or the 
CERN Large Hadron Collider (LHC) energies, requires a sizable admixture of binary collisions~\cite{Back:2001xy,Kharzeev:2000ph}, 
the scaling based on wounded quarks~\cite{Bialas:1977en,Bialas:1977xp,Bialas:1978ze,Anisovich:1977av} works remarkably 
well~\cite{Eremin:2003qn,KumarNetrakanti:2004ym,Bialas:2006kw,Bialas:2007eg,Alver:2008aq,Agakishiev:2011eq,Adler:2013aqf,Loizides:2014vua,Adare:2015bua,%
Lacey:2016hqy,Bozek:2016kpf,Zheng:2016nxx,Sarkisyan:2016dzo,Mitchell:2016jio,Chaturvedi:2016ctn,Loizides:2016djv,Tannenbaum:2017ixt}. Another 
successful approach~\cite{Zakharov:2016bob,Zakharov:2016gyu} amends the wounded 
nucleons with a meson-cloud component. 

For mid-rapidity production, the wounded quark scaling takes the simple form
\begin{eqnarray}
N_{\rm ch} = k (\br{N_A}+\br{N_B}), \label{eq:wounded}
\end{eqnarray}
where $N_{\rm ch}$ is the number of charged hadrons in a mid-rapidity bin, and $\br{N_i}$ are the average numbers of wounded quarks in 
nucleus $i$ in a considered centrality class. 
The proportionality constant $k$ should not depend on centrality or the mass numbers of the nuclei (i.e., on the overall number of participants), 
and indeed this requirement is satisfied to expected accuracy~\cite{Bozek:2016kpf,Tannenbaum:2017ixt}. Of course, $k$ increases with the collision energy. 

When it comes to modeling the rapidity spectra, formula~(\ref{eq:wounded}) is replaced with 
\begin{eqnarray}
\frac{dN}{d\eta} = \br{N_A} f(\eta)+\br{N_B} f(-\eta), \label{eq:woundeta}
\end{eqnarray}
where $f(\eta)$ is a universal (at a given collision energy) profile for emission from a wounded quark
(we adopt the convention that nucleus $A$ moves to the right and $B$ to the left). 
For symmetric ($A=B$) collisions one only gets access to the the symmetric part of $f(\eta)$, as then 
$\br{N_A}=\br{N_B}$. However, from asymmetric collisions, such as d-Au, one can also extract
the antisymmetric component in the wounded nucleon~\cite{Bialas:2004su} or wounded quark model~\cite{Barej:2017kcw,Adare:2018toe}
(for A-A collisions analogous analyses were carried out in~\cite{Gazdzicki:2005rr,Bzdak:2009dr,Bzdak:2009xq}), 
with the finding that $f(\eta)$ is peaked
in the forward region, thus quite naturally emission is in the forward direction. However, $f(\eta)$ is widely spread in the whole kinematically available range. 
The phenomenological result of the approximate triangular shape of the emission profile was later used in modeling the initial 
conditions for further evolution, see e.g.~\cite{Adil:2005bb,Bozek:2010bi,Bozek:2013uha,Monnai:2015sca,Chatterjee:2017mhc}.

Microscopically, the approximate triangular shape of the emission function finds a natural origin in color string models, where 
one end-point of the string is fixed, whereas the location of the other end-point fluctuates. 
In particular, 
in the basic Brodsky-Gunion-Kuhn mechanism~\cite{Brodsky:1977de}, the emission proceeds from strings in which 
one end-point is associated with a valence parton, and the other end-point, corresponding to 
wee partons, is randomly generated along the space-time rapidity $\eta$. When the distribution of the fluctuating end-point is 
uniform in $\eta$, and so is the string fragmentation distribution, then the triangular 
shape for the emission function follows. 

Various Monte Carlo codes implementing the Lund string formation and 
decays (see, e.g.,~\cite{Andersson:1983ia,Wang:1991hta,Lin:2004en,Sjostrand:2014zea,Bierlich:2018xfw,Ferreres-Sole:2018vgo}) or the dual parton 
model/Regge-exchange approach~\cite{Capella:1992yb,Werner:2010aa,Pierog:2013ria}
also introduce strings of fluctuating ends, with various specific mechanisms and effects (baryon stopping, nuclear shadowing) additionally incorporated. 
Apart from reproducing the measured one-body spectra, achieved by appropriate tune-ups of parameters, the incorporated initial-state  
correlations show up
in event-by-event fluctuations that can be accessed experimentally.
Thus the fluctuating strings  are standard objects used in modeling the early phase of high-energy reactions.

Our model joins the concept of wounded sources with strings in the following way:
\begin{enumerate}
 \item Each wounded source has an associated string.
 \item The strings emit particles independently of each-other.
 \item The end-points of a string are generated universally (in the same manner for all wounded objects) from appropriate distributions.
 \item The emission of particles from a string occurring between the end-points is homogeneous in spatial rapidity.
\end{enumerate}
In such a model, event-by-event fluctuations take 
the origin from fluctuations of the number of  wounded objects, as well as from fluctuations of the positions of the end points~\cite{Broniowski:2015oif}. 
The goal of this paper is to study this generic model, with the focus on the end-point behavior which probes the underlying physics. 
We take a general approach, with no prejudice as to how the end-points are fluctuating, but using the one-body emission profiles 
obtained from experiment as a physical constraint. 

More complicated mechanisms associated with dense systems,
such as the formation of color ropes~\cite{Biro:1984cf,Sorge:1995dp} or nuclear shadowing, are not incorporated in our picture. Also, we consider 
one type of strings, which allows for simple analytic derivations.  

We remark that associating a string with a leading quark is in the spirit of the Lund approach (cf. discussion of Sec.~5 in~\cite{Andersson:1983ia}). 
So for simplicity we have in each event $N_i$ ``wounded strings'' associated with valence quarks in nucleus $i$. Other more complicated choices
(e.g, including the binary collisions) are also 
possible here, but the advantage of our prescription is that by definition it complies with the experimental scaling of multiplicities of Eq.~(\ref{eq:wounded}).

A specific implementation of some ideas explored in this work, with strings that have one end fixed and the other 
fluctuating, has been presented in~\cite{Broniowski:2015oif}. 

The outline of our paper is as follows:

In Sec.~\ref{sec:wound} we use the rapidity spectra from d-Au and Au-Au reactions 
at $\sqrt{s_{NN}}=200$~GeV to obtain the  one-body emission profile of the wounded quark.
In Sec.~\ref{sec:model} we explore our generic string model and derive 
simple relations between string end-point distributions and n-body-emission profiles for the radiation from 
individual strings. 
Section~\ref{sec:end} discusses how a given one-body-emission profile can correspond to a 
family of different functions for the string end-point distributions. 
Two-body correlations from a single string are discussed in Sec.~\ref{sec:corr}, whereas in
Sec.~\ref{sec:corrm} they are combined to form the two-body correlations in nuclear collisions.
Section~\ref{sec:anm} presents the Legendre $a_{nm}$ coefficients of the two-particle correlations. Finally,  Sec.~\ref{sec:concl} draws the final conclusions from our work.
Some more technical developments can be found in the appendices.

\section{Emission profiles from wounded quarks \label{sec:wound}}

\begin{figure}
\hfill \includegraphics[trim={22pt 0 0 0},clip,width=0.48\textwidth]{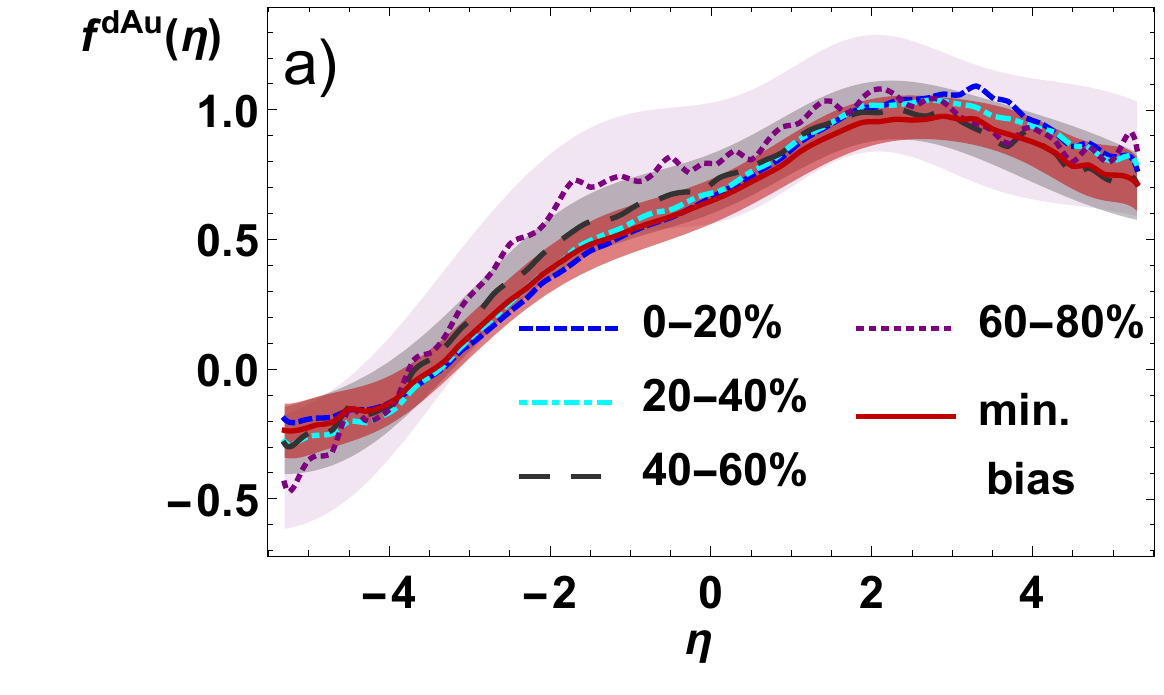}\\
\hfill \includegraphics[trim={22pt 0 0 0},clip,width=0.47\textwidth]{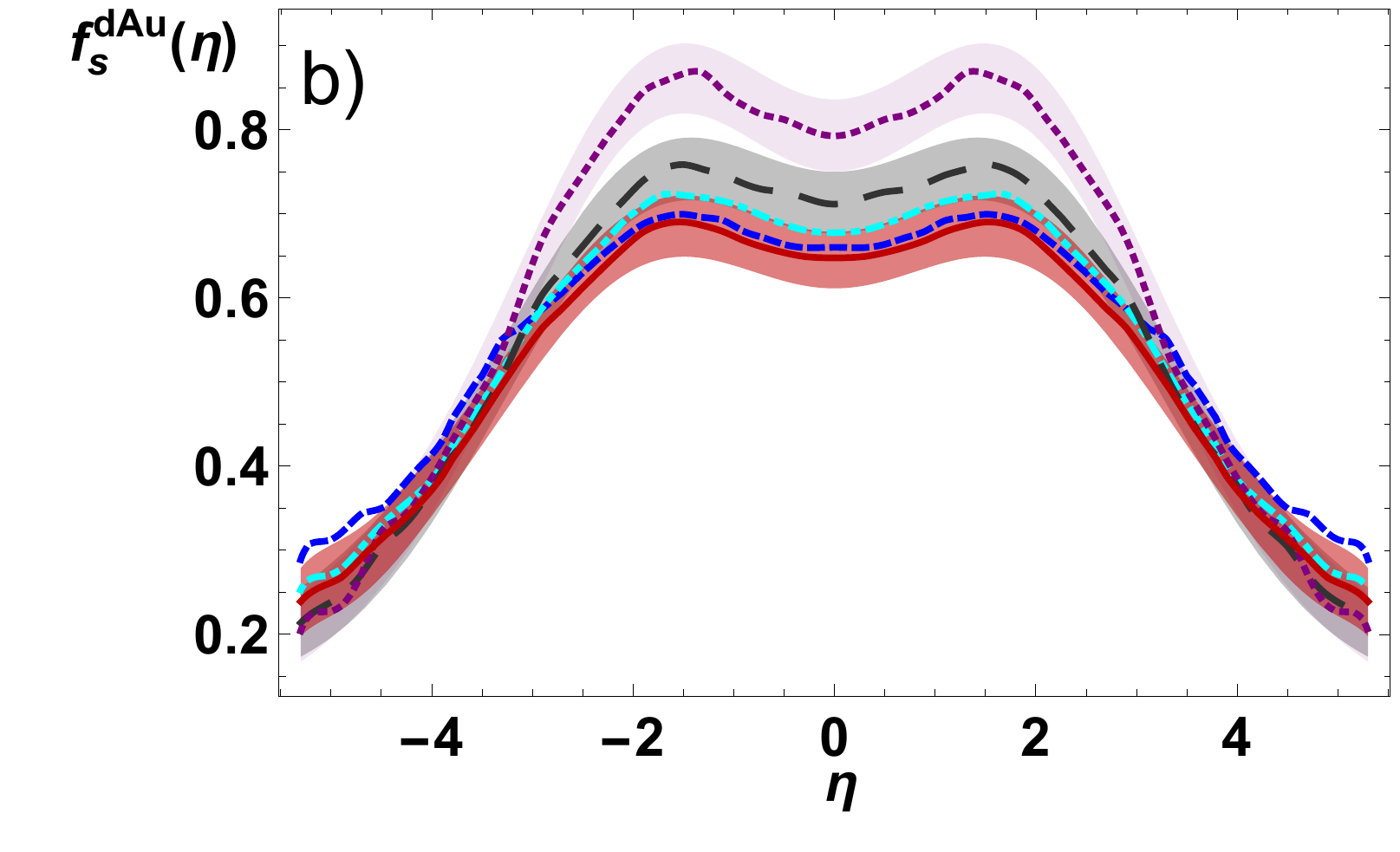}\\
\hfill \includegraphics[trim={26pt 0 0 0},clip,width=0.48\textwidth]{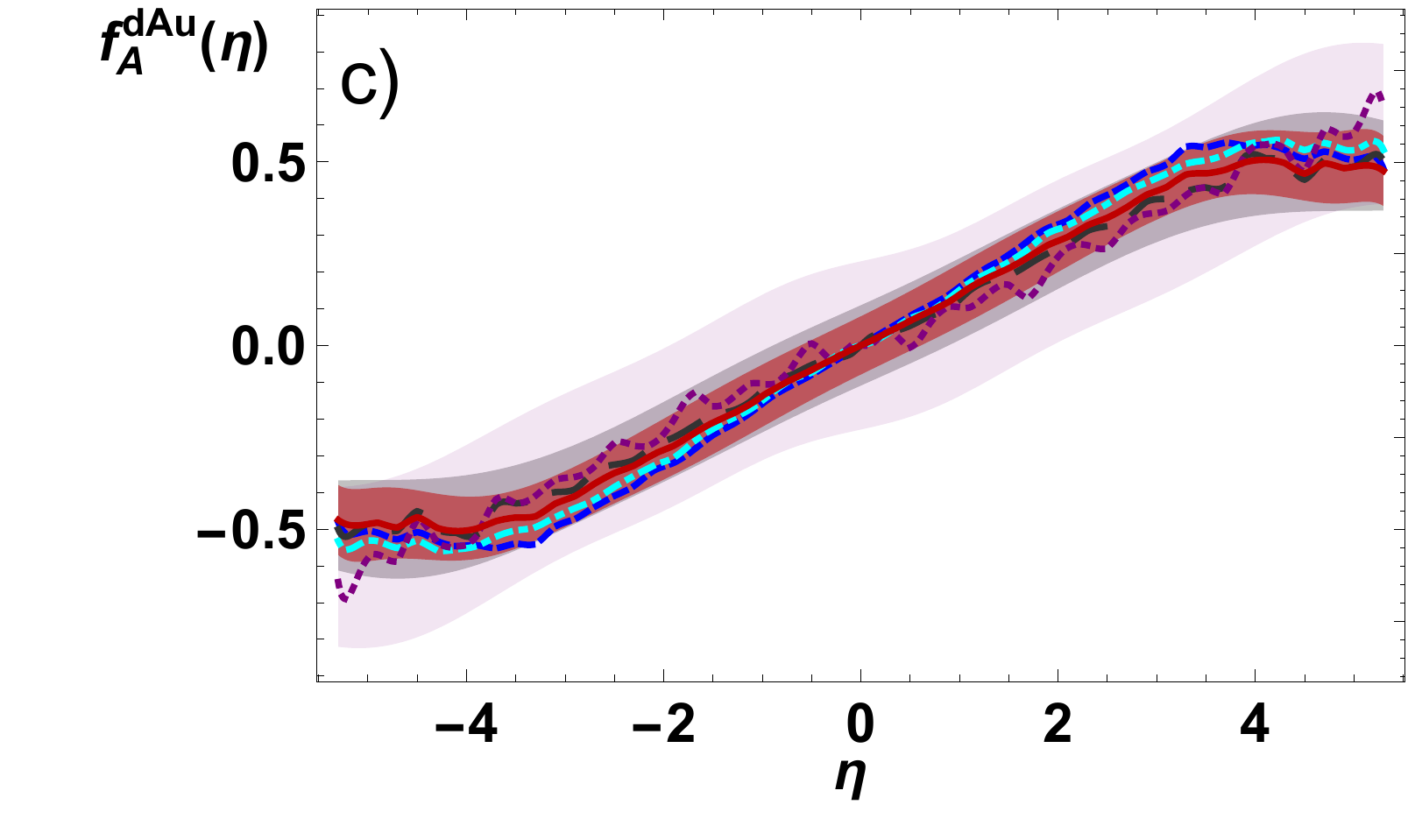}\\
\vspace{-5mm}
\caption{One-particle emission profiles obtained in the wounded quark model via Eqs.~(\ref{eq:woundeta}-\ref{eq:s_a_wound}) from the PHOBOS
rapidity spectra for d-Au collisions at $\sqrt{s_{NN}}=200$~GeV~\cite{Back:2004mr} in the indicated centrality classes (a), together with 
the corresponding symmetric (b) and antisymmetric (c) components. 
The shaded bands show the experimental uncertainties (propagated via the Gaussian method) for 
the $40-60$\% and $60-80$\% centrality classes, as well as for the PHOBOS minimum bias data~\cite{Back:2003hx}. \label{fig:dAu}}
\end{figure}

We begin by obtaining from experimental data the emission profiles of Eq.~(\ref{eq:woundeta}), needed in the following sections.  
We use the method of~\cite{Bialas:2004su}, which  has also been  
applied recently to wounded quarks in~\cite{Barej:2017kcw}. With 
\begin{eqnarray}
&& f_s(\eta) = \frac{1}{2}[f(\eta)+f(-\eta)], \;\;   f_a(\eta) = \frac{1}{2}[f(\eta)-f(-\eta)], \nonumber  \\
&& N_+=N_A+N_B, \;\; N_-=N_A-N_B, 
\end{eqnarray}
one gets immediately
\begin{eqnarray}
&& f_s(\eta)=\frac{dN/d\eta(\eta)+dN/d\eta(-\eta)}{\br{N_+}}, \nonumber \\ 
&& f_a(\eta)=\frac{dN/d\eta(\eta)-dN/d\eta(-\eta)}{\br{N_-}}.
\label{eq:s_a_wound}
\end{eqnarray}
For asymmetric collisions both parts of the profile can be obtained, whereas for symmetric collisions one can only get $f_s(\eta)$.

\begin{figure}
\begin{center}
\includegraphics[trim={10pt 0 0 0},clip,width=0.48\textwidth]{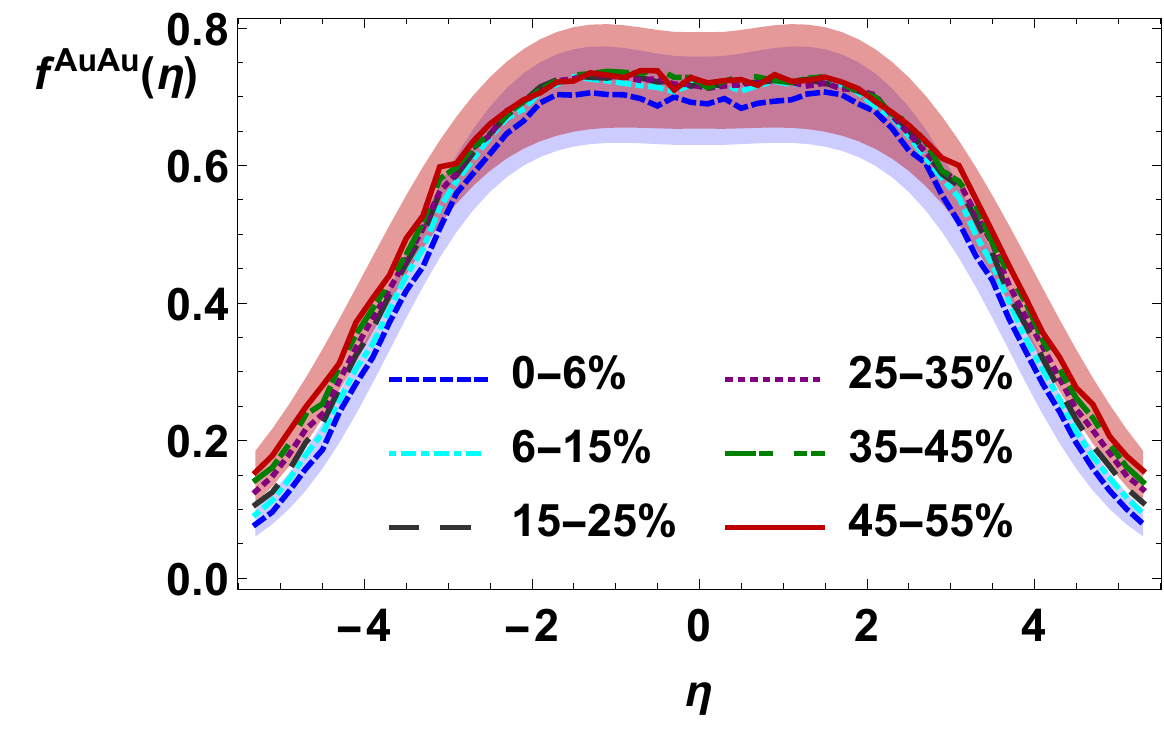}
\end{center}
\vspace{-7mm}
\caption{One-particle emission profiles obtained in the wounded quark model from the PHOBOS rapidity 
spectra for Au-Au collisions at $\sqrt{s_{NN}}=200$~GeV~\cite{Back:2002wb} in the indicated centrality classes. The shaded bands give the experimental uncertainties 
(propagated via the Gaussian method) for the most central and the most peripheral case. \label{fig:AuAu}}
\end{figure}

\begin{figure}
\hfill \includegraphics[trim={2pt 0 0 0},clip,width=0.48\textwidth]{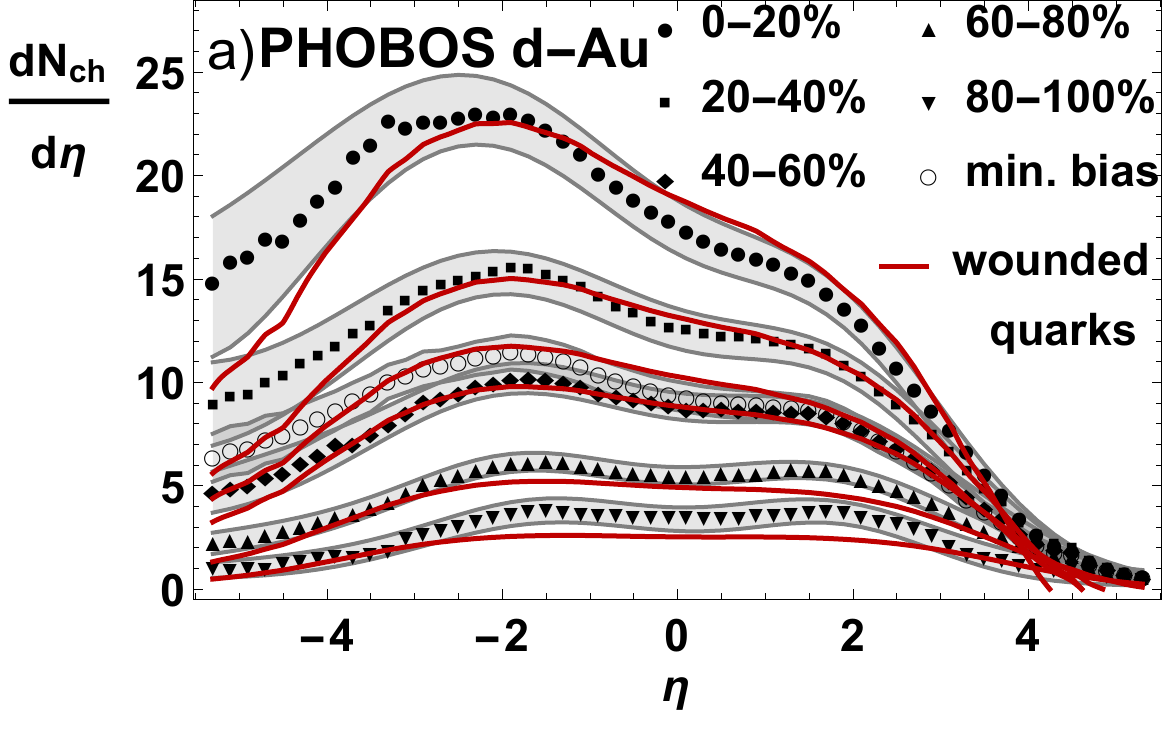}\\
\hfill \includegraphics[trim={10pt 0 0 0},clip,width=0.482\textwidth]{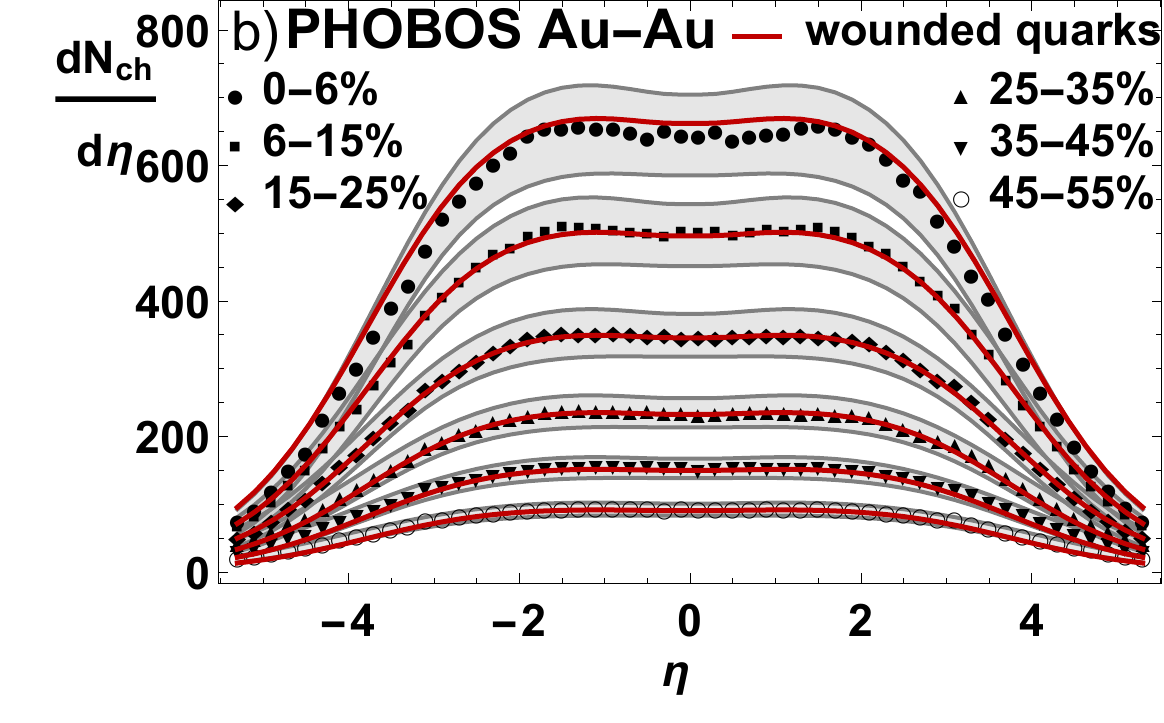}\\
\vspace{-5mm}
\caption{
Comparison of the wounded-quark model predictions (lines) with the experimental rapidity spectra (points) for d-Au~\cite{Back:2004mr} (a) and Au-Au~\cite{Back:2002wb} (b) collisions,  
with the experimental uncertainties shown as shaded bands. A universal profile discussed in the text is taken for the model calculations in all cases. 
\label{fig:dAu+AuAu}}
\end{figure}

If the wounded-quark scaling works, then the profiles obtained with different centrality classes or mass numbers of the colliding nuclei should be universal, depending 
only on the collision energy. To what extent this is the case, can be assessed 
from Figs.~\ref{fig:dAu} and \ref{fig:AuAu}, which show the one-particle emission 
profiles that were extracted from experimental data on d-Au and Au-Au collisions from the
PHOBOS data~\cite{Back:2003hx,Back:2004mr,Back:2002wb} in the framework of the wounded quark model. 
To this end, the symmetric (for both reactions) and antisymmetric components (only in the case of the d-Au 
collisions) were obtained from the experimental data on rapidity spectra by means of Eq.~(\ref{eq:s_a_wound}), 
where the valence quark multiplicities $\br{N_\pm}$ were obtained from 
{\tt GLISSANDO}~\cite{Broniowski:2007nz,Rybczynski:2013yba}, a Monte-Carlo simulator of the Glauber model.

Figure~\ref{fig:dAu} shows the results for the one-particle-emission profiles $f^{\rm dAu}(\eta)$ extracted from the PHOBOS data~\cite{Back:2003hx,Back:2004mr} 
for d-Au collisions, together with their symmetric and antisymmetric components. 
In general, the curves for various centrality classes, considering the propagated experimental errors, can be viewed as coinciding.
The apparent exception to this behavior is seen in the symmetric part of the profile for the peripheral centrality $60\%-80\%$, which is 
significantly larger for $|\eta|<3$, cf. Fig.~\ref{fig:dAu}(b). We note that for d-Au collisions 
this peripheral class corresponds to $\langle N_+\rangle$ in the range from six to eight sources, which are tiny values, where the model admittedly does not work.
It can thus confirm the 
findings of~\cite{Barej:2017kcw} that the assumption of universality of the one-particle emission profiles works reasonably well for the central to mid-peripheral d-Au
collisions, whereas it starts to differ for more peripheral centrality classes.

Figure~\ref{fig:AuAu} presents our results for the one-particle emission profiles $f^{\rm AuAu}(\eta)$ extracted from the PHOBOS data~\cite{Back:2002wb} 
for Au-Au collisions. As already mentioned, in this case only the symmetric parts of  the emission profiles can be obtained. 
It can be seen that the results for $f^{\rm AuAu}(\eta)$ 
in various centrality classes agree remarkably well with one another. They also approximately agree with the symmetric profiles for d-Au collisions
of Fig.~\ref{fig:dAu}(b).

Finally, we test if our method  reproduces the PHOBOS charged particle rapidity spectra
for combined d-Au and Au-Au collisions. To this end we take a single ``universal'' $f(\eta)$, consisting 
of an antisymmetric part extracted from the minimum-bias d-Au spectra and a symmetric part  taken as 
the average of the different one-particle emission profiles of Au-Au collisions shown in Fig.~\ref{fig:AuAu}.  
The charged particle rapidity spectra $dN_{\rm ch}/d\eta$ were calculated by means of Eq.~(\ref{eq:woundeta}) with this universal $f(\eta)$, 
where again the numbers $\br{N_A}$ and $\br{N_B}$ were generated with {\tt GLISSANDO}.
Figure~\ref{fig:dAu+AuAu} shows the resulting one-particle-emission spectra for d-Au and Au-Au collisions obtained that way,
together with the corresponding experimental data from PHOBOS~\cite{Back:2003hx,Back:2004mr,Back:2002wb}:
As expected from Fig.~\ref{fig:AuAu}, the rapidity spectra for the Au-Au collisions, which are almost symmetric, are very well reproduced by the chosen $f(\eta)$.
Also the rapidity spectra for the d-Au collisions, which largely depend on both the symmetric and antisymmetric contribution to $f(\eta)$, are qualitatively well reproduced for 
$|\eta|<4$, except for the above-discussed case of the peripheral collisions.

Therefore, we conclude that the wounded quark model with the universal profile function $f(\eta)$ reproduces the experimental rapidity spectra 
at $\sqrt{s_{NN}}=200$~GeV
in a way satisfactory for our exploratory study.\footnote{We note that the analogous analysis at the LHC leads to somewhat less accurate agreement, which calls for 
improvement of the model.}
In the following analysis of the rapidity fluctuations, we use the $f(\eta)$ obtained here to constrain the string end-point distributions.

\section{Generic string model \label{sec:model}}

In this section we describe a model of generic production from a single string formed in the early phase of the collision process. 
Suppose the string is pulled by two end-points placed at spatial rapidities $y_{1}$ and $y_{2}$, whose locations are generated according to a probability 
distribution $g(y_1,y_2)$ (if the end-points are generated in an uncorrelated manner, then $g(y_1,y_2)=g_1(y_1)g_2(y_2)$, as will
be assumed shortly). The emission of a particle with rapidity $\eta$ from the string fragmentation process 
is assumed to be uniformly distributed along the string, i.e., it is equal to
\begin{eqnarray}
\!\!\!\!\! s(\eta;y_1,y_2)= \omega \left [ \theta(y_1<\eta<y_2) + \theta(y_2<\eta<y_1) \right ], \label{eq:uni}
\end{eqnarray}
where $\omega$ is a dimensionless constant determining the production strength and $\theta(c)$ imposes the condition $c$. 
Note that we include the cases of $y_2>y_1$ and $y_1>y_2$, which 
may seem redundant but which is needed, for instance, when the two end-points correspond to different partons in a given model.  

Let us introduce the short-hand notation 
\begin{eqnarray}
\int_{\cal Y} dy_1 dy_2 \,  g(y_1,y_2) X = \langle X \rangle_{\cal Y},
\end{eqnarray}
with ${\cal Y}$ denoting the two-dimensional range of integration, depending on the kinematic constraints  and/or detector coverage, and $X$ meaning
any expression.
The single-particle density for production from a string upon averaging over the fluctuation of the end-points is therefore 
\begin{eqnarray}
f(\eta) = \langle s(\eta;y_1,y_2) \rangle_{\cal Y} ,
\label{eq:1}
\end{eqnarray}

Analogously, for the $n$-particle production ($n\ge 2$)  from a single string we have
\begin{eqnarray}
f_n(\eta_1,\dots,\eta_n) =\langle  s(\eta_1;y_1,y_2) \dots s(\eta_n;y_1,y_2) \rangle_{\cal Y}, \label{eq:n}
\end{eqnarray}
where we have assumed independent production of the $n$ particles. 

In case the string ends are generated independently of each other, one has 
\begin{eqnarray}
\langle X \rangle_{\cal Y} =  \int dy_1 dy_2  g_1(y_1)g_2(y_2) X\, ,
\end{eqnarray}
where the limits of integration in $y_i$ are formally from $-\infty$ to $\infty$, with the support taken care of by the forms 
of $g_{i}(y_i)$. Then we readily find that the one-body emission profile is
\begin{eqnarray}
f(\eta) &=& \omega \left \{ G_1(\eta)[1-G_2(\eta)] + G_2(\eta)[1-G_1(\eta)] \right \} \nonumber \\ 
&=& \omega \left \{ \tfrac{1}{2} - 2 [G_1(\eta)-\tfrac{1}{2}]   [G_2(\eta)-\tfrac{1}{2}]    \right \}, \label{eq:f1G} 
\end{eqnarray}
where the appropriate cumulative distribution functions (CDFs) are defined as 
\begin{eqnarray}
G_{i}(y)=\int_{-\infty}^y dy' \, g_{i}(y') . 
\end{eqnarray}

The profile $f(\eta)$ acquires a specific value at the arguments $\eta_1$ and $\eta_2$ where the CDFs reach $\tfrac{1}{2}$, i.e.,  
\begin{eqnarray}
&& \eta_1^{(0)} \; : \; G_1(\eta_1^{(0)})=\tfrac{1}{2}, \nonumber \\  
&& \eta_2^{(0)} \; : \; G_2(\eta_2^{(0)})=\tfrac{1}{2}. \label{eq:e1e2}
\end{eqnarray}
Then from Eq.~(\ref{eq:f1G}) we obtain
\begin{eqnarray}
\omega=2 f(\eta_1^{(0)}) =  2 f(\eta_2^{(0)}). \label{eq:1o2}
\end{eqnarray}
This equation provides a special meaning to the constant $\omega$.
Furthermore, since $0 \le G_{1,2}(\eta) \le 1$, Eq.~(\ref{eq:f1G}) yields the limit
\begin{eqnarray}
0 \le f(\eta)\le \omega. 
\label{eq:fupbound}
\end{eqnarray}
The above features will be explored shortly in a qualitative discussion.

Similarly, for the $n$-particle distributions with $n\ge2$ we have
\begin{eqnarray}
&& f_n(\eta_1,\dots,\eta_n) = \omega^n \left \{ \right. \label{eq:fnG} \\ 
&& ~~G_1({\rm min}(\eta_1,\dots,\eta_n))[1-G_2({\rm max}(\eta_1,\dots,\eta_n))] + \nonumber \\
&& \left . ~~G_2({\rm min}(\eta_1,\dots,\eta_n))[1-G_1({\rm max}(\eta_1,\dots,\eta_n))] \right \}. \nonumber 
\end{eqnarray}

We thus see that in the model with two end-points fluctuating (the relevant assumptions are the uniform string fragmentation (\ref{eq:uni}) and the independence 
of the two end-point locations)  
all the information carried by the $n$-particle densities produced from a single string is encoded solely in the cumulative distributions 
functions $G_1$ and $G_2$. 
It is obvious, however, that $G_1$ and $G_2$ cannot be  separately determined from the one-body distributions in an unambiguous manner, 
hence a large degree of freedom is still left in the model after fixing the rapidity spectra. Yet, the one body distribution provides, via Eq.~(\ref{eq:f1G}), an
important constraint. Our method of matching  $G_1$ and $G_2$ to the one-body function $f(\eta)$ is explained in detail in Appendix~\ref{app:match}.
As we stress, there is no uniqueness in the procedure, but there is a systematic way of approaching the problem, allowing one to explore the range of possibilities.  

We denote the position of the maximum of $f(\eta)$ as $\eta_{\rm max}$. 

We consider three cases:
\begin{enumerate}
 \item[i)]  The distributions of both end-points are equal, $g_1(\eta)=g_2(\eta)$, Eq.~(\ref{eq:case1}).
 In this case $\omega=2f(\eta_{\rm max})$, with $\eta_{\rm max}=\eta_1^{(0)}=\eta_2^{(0)}$.
 \item[ii)] The supports of  distributions $g_1(\eta)$ and $g_2(\eta)$ do not overlap, Eq.~(\ref{eq:case2}). 
In this case $\omega=f(\eta_{\rm max})$ and  $\eta_2^{(0)} < \eta_{\rm max}<\eta_1^{(0)}$.
 \item[iii)]  The form of $g_1(\eta)$  is motivated by parton distribution functions (PDFs) of valence quarks, Eq.~(\ref{eq:pdffit}), and $g_2(\eta)$
   is adjusted according to Eq.~(\ref{eq:h2}).
\end{enumerate}
Cases i) and ii) are in a sense most different, showing the span of possibilities formally allowed, whereas case iii) is intermediate. 
For case iii) we use the parametrization of the valence quark PDF given by Eq.~(\ref{eq:pdffit}) with parameters 
$\alpha=-0.5$ and $\beta=3$, which are typical values at low scales.
We have found that using other reasonable parametrizations has very small influence on our results, with case iii) always remaining
close to case i).

We stress that all the considered cases reproduce, by construction, the one-body emission profiles $f(\eta)$. 

We end this section with remarks concerning the model with one end of the string fixed  and the other one fluctuating, explored in~\cite{Broniowski:2015oif}.
This simplified version 
can be obtained as a special limit from Eqs.~(\ref{eq:f1G},\ref{eq:fnG})
by choosing \mbox{$g_1(\eta)=\delta(\eta-y_{\rm max})$}, which is equivalent of taking, correspondingly, $G_1=0$ 
for $\eta < y_{\rm max}$, {\em i.e.},
\begin{eqnarray}
f(\eta) &=& \omega  G_2(\eta),  \label{eq:f1Gsingle} \\
f_n(\eta_1,\dots,\eta_n) &=& \omega^n  G_2({\rm min}(\eta_1,\dots,\eta_n)). \nonumber
\end{eqnarray}
We note immediately that this model cannot reproduce $f(\eta)$ for $\eta > \eta_{\rm max}$, as $G_2(\eta)$ cannot decrease.
Thus the model is limited to   $\eta \le \eta_{\rm max}$, which, however, is not a problem if we are only interested in the mid-rapidity region. 

Moreover, in this region the single-end fluctuating model corresponds precisely to case ii) of the two-end fluctuations. This is obvious from the 
following argumentation: When the right end of the string is fluctuating outside of the acceptance region, it is irrelevant if it fluctuates or if it is 
fixed, as in both cases we only observe the production from the part of the string falling into the acceptance range. In that situation
(or more precisely for  $\eta \le \eta_{\rm max}$) Eqs.~(\ref{eq:f1G},\ref{eq:fnG}) reduce to Eqs.~(\ref{eq:f1Gsingle}). Hence,  the single
end-point fluctuation model of~\cite{Broniowski:2015oif} corresponds to the present case ii) at   $\eta \le \eta_{\rm max}$, and is not applicable for 
$\eta > \eta_{\rm max}$.

\section{End-point distributions \label{sec:end}}

We now come to the discussion of the end-point distributions subjected to the requirement that the one-body 
emission profiles are reproduced.

\begin{figure}
\hfill \includegraphics[trim={11pt 0 1pt 0},clip,width=0.49\textwidth]{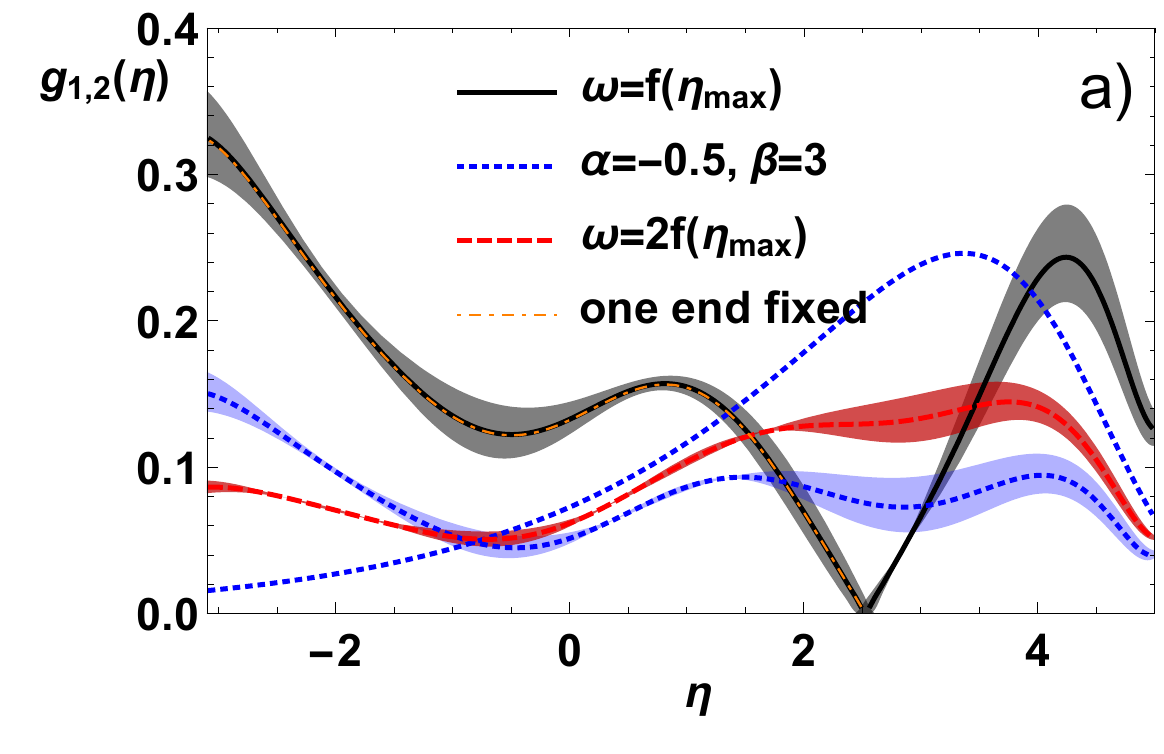}\\ \vspace{-3mm}
\hfill \includegraphics[trim={13pt 0 1pt 0},clip,width=0.49\textwidth]{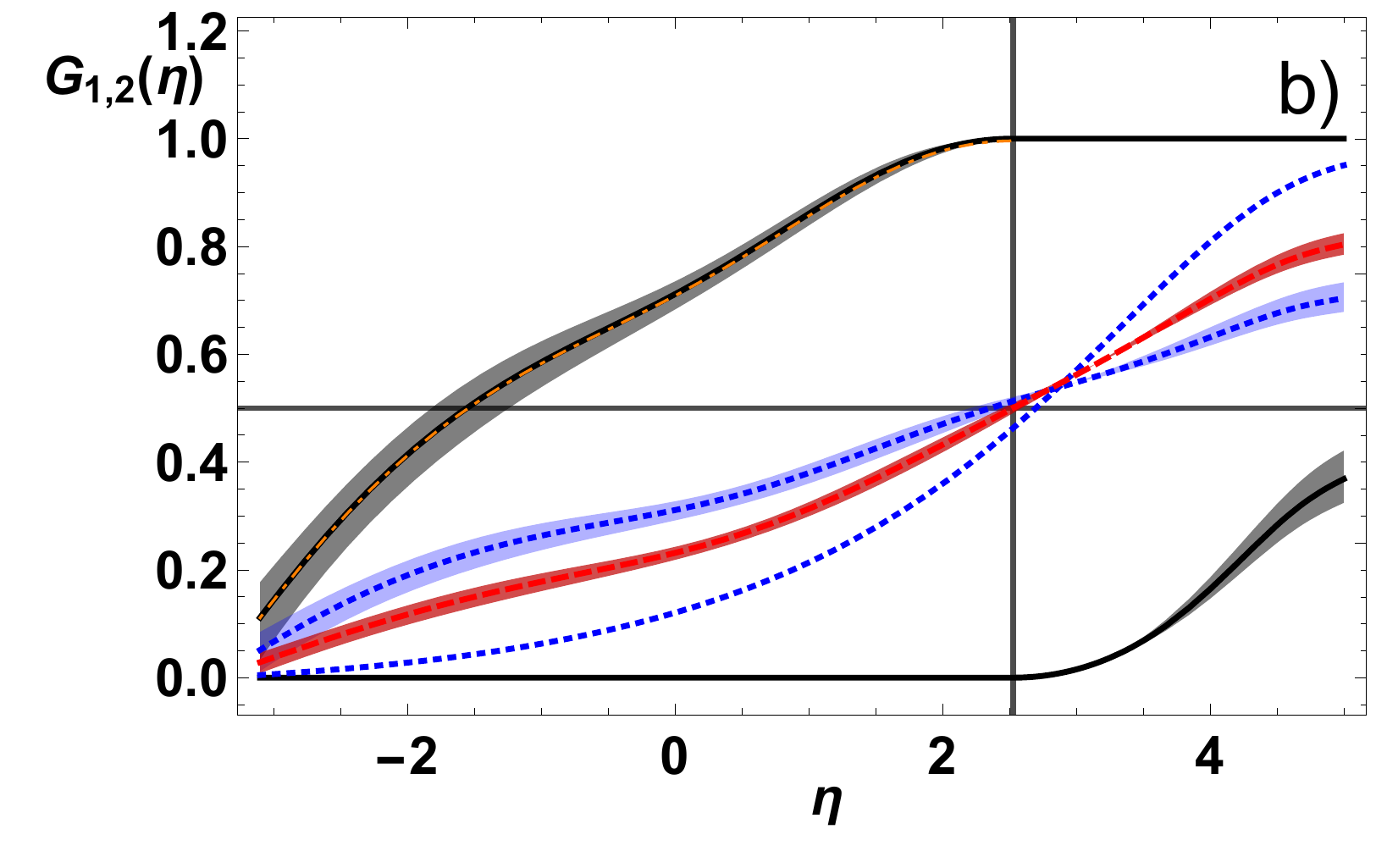}\\
\vspace{-5mm}
\caption{Distribution functions $g_1$ and $g_2$ (a) and cumulative distribution functions $G_1$ and $G_2$ 
(b)  of the string end-points for the cases of i) $\omega=2f(\eta_{\rm max})$, ii)~$\omega=f(\eta_{\rm max})$, and iii)~$\alpha=-0.5,$ $\beta=3$, 
as indicated in the legend. The light dot-dashed line corresponds to the model with one end-point fixed and the other one fluctuating~\cite{Broniowski:2015oif},
which overlaps with case ii) for $\eta \le \eta_{\rm max}$. The vertical line in panel (b) indicates $\eta=\eta_{\rm max}$. 
See the text for further details. \label{fig:g1g2}}
\end{figure}

Figure~\ref{fig:g1g2}a) shows the distributions of the string end-points, $g_1$ and $g_2$, for the three cases, and Fig.~\ref{fig:g1g2}b) the corresponding CDFs, $G_1$ and $G_2$. 
The shaded bands give an estimate of the errors due to the experimental uncertainty $\Delta f$ for the one-particle emission profile $f$. 
In the case of Fig.~\ref{fig:g1g2}b), the upper limit of the shaded bands corresponds to the values of $G_{1,2}$ that are matched to the 
one-body profile $f+\Delta f$, whereas the lower limits are matched to $f-\Delta f$. 
For these upper and lower limits of $G_{1,2}$, the derivatives in $\eta$ yield the upper and lower limits of the shaded bands 
for $g_1$ and $g_2$ depicted in Fig.~\ref{fig:g1g2}a).
For case iii) a shaded band is given only for $g_2$  ($G_2$). This is because by construction $g_1$  ($G_1$) coming from PDFs are 
assumed to be accurate and all uncertainty is therefore attributed to $g_2$ ($G_2$).

In case i) $g_1(\eta)=g_2(\eta)$, hence the distributions are indicated with a single curve (solid line) in Figs.~\ref{fig:g1g2}a) and b). 
We note that the distribution of $g_{1}(\eta)$ peaks at forward rapidity (the Au side), 
as expected from the shape of the one-body profile $f(\eta)$ in Fig.~\ref{fig:dAu}. The CDF crosses 
the value $1/2$ at $\eta_1^{(0)}=\eta_2^{(0)}=\eta_{\rm max}\simeq 2.5$, 
which coincides with the maximum of $f(\eta)$.

In case ii) (dashed lines in Fig.~\ref{fig:g1g2}) the supports for $g_1$ and $g_2$ are disjoint. In Fig.~\ref{fig:g1g2}a) the left part of the curve, up to the 
point $\eta_{\rm max} \simeq 2.5$ (indicated with a vertical line), corresponds to $g_2$, and the right part to $g_1$. Hence,  
the string end-points always follow the ordering $y_1 \ge y_2$, 
which does not hold in the other cases. 
Figure~\ref{fig:g1g2}b) shows the corresponding CDFs, with $G_1=0$ left from $\eta_{\rm max}$, and  $G_2=1$ right from $\eta_{\rm max}$.
In Appendix~\ref{app:match}
we show that $G_1$ and $G_2$ from case ii) are the lower and upper limits for any CDFs in the considered problem. Indeed, the CDFs from the other two cases 
fall in between these limiting curves. 

Case iii), based on a valence quark PDF for $g_1$, represents an intermediate class of distributions falling between 
cases i) and ii). The curves corresponding to the valence quark are dotted and with no error bands. The 
distribution $g_1$ (valence quark) is peaked in the forward direction, as expected. 
We note that $y_1>y_2$ is favored, although $y_2<y_1$ is also possible. 
With the parametrization we used of the valence quark distribution, the CDFs in case iii) are not far from case i).
We have checked that this feature holds also for other reasonable parametrizations of the valence quark PDF.

We underline again that all the cases of Fig.~\ref{fig:g1g2}, which exhibit radically different end-point distributions, reproduce by construction the
one-body emission profile $f(\eta)$.

\section{Correlations from a single string \label{sec:corr}}

\begin{figure}
\begin{center}
\includegraphics[width=0.48\textwidth]{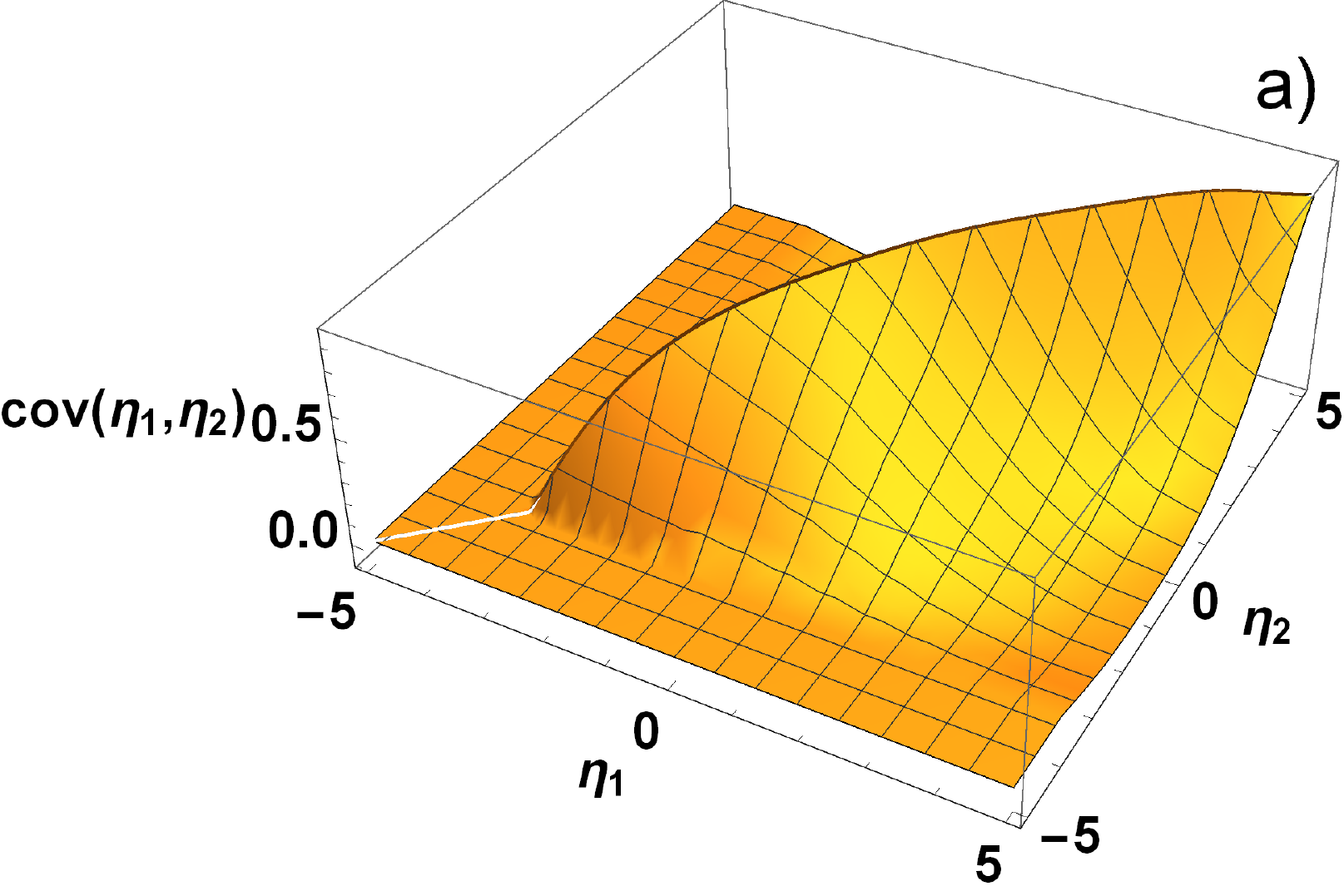}
\includegraphics[width=0.48\textwidth]{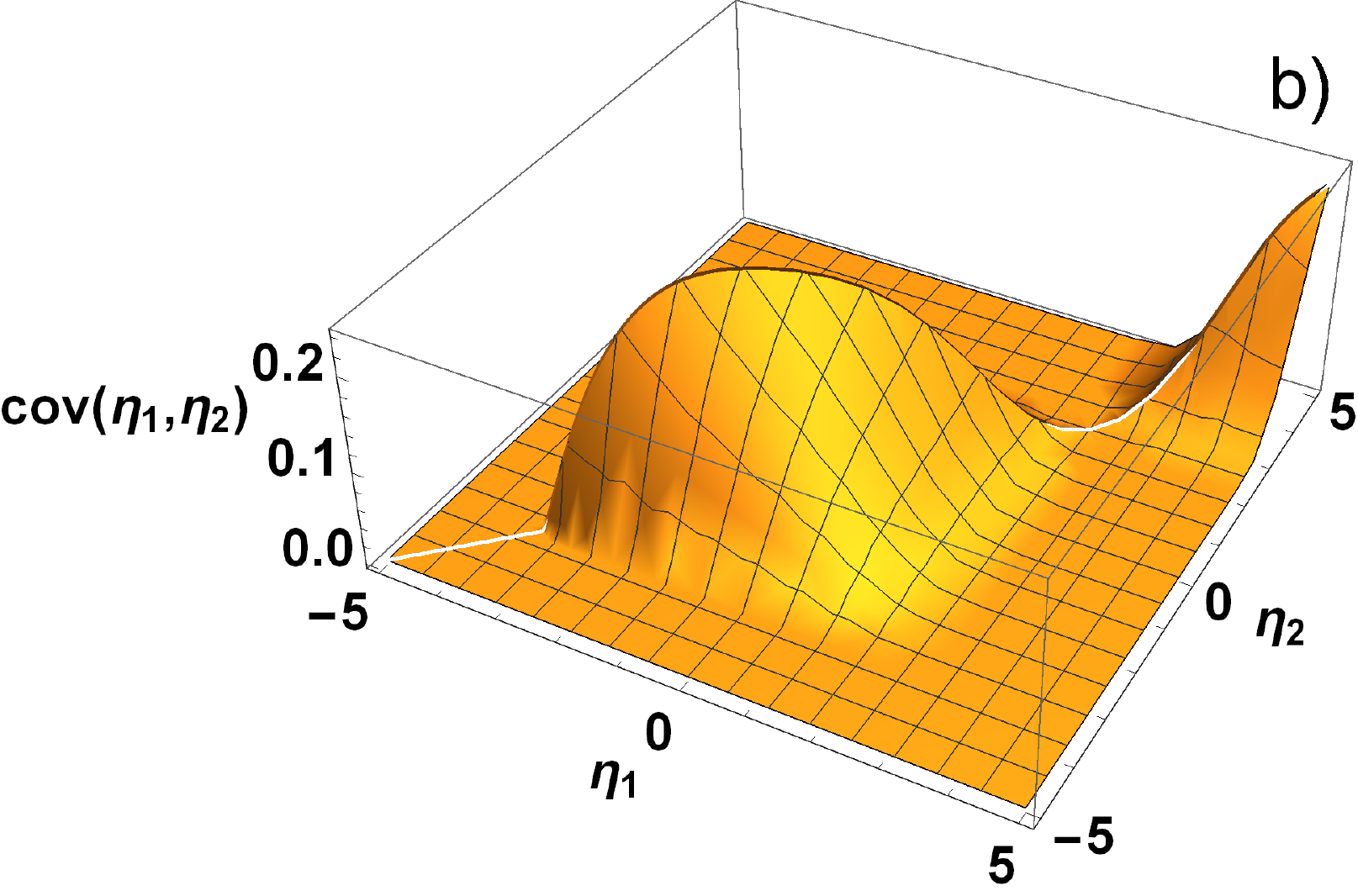}
\includegraphics[width=0.48\textwidth]{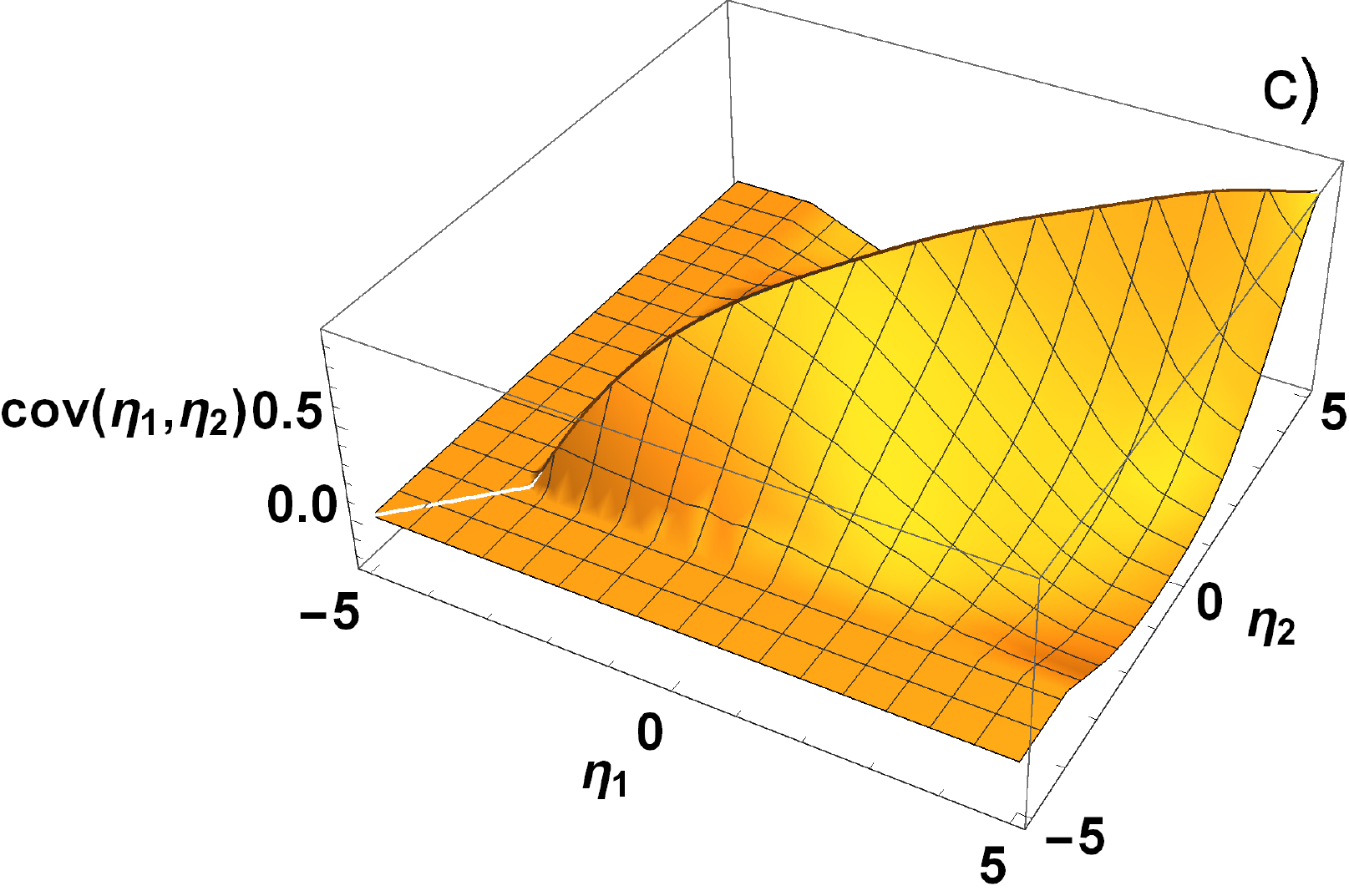}
\end{center}
\vspace{-5mm}
\caption{Covariance for the emission from a single string for cases i) (a), ii) (b) and iii) (c). \label{fig:cov}}
\end{figure}

As we show in this section, the two-particle correlation is sensitive to the particular form of the distributions and differs between cases i), ii), and iii). 
A convenient quantity is the covariance of the two-particle emission
from a single string, defined as
\begin{eqnarray}
{\rm cov}(\eta_1,\eta_2)={f_2(\eta_1,\eta_2)} -  {f(\eta_1)} {f(\eta_2)}, \label{eq:cov} 
\end{eqnarray}
where $f_2$ is given by Eq.~(\ref{eq:fnG}).
Explicitly, 
\begin{eqnarray}
&& {\rm cov}(\eta_1,\eta_2)=\omega^2 \big \{ G_1({\rm min}(\eta_1,\eta_2))[1-G_2({\rm max}(\eta_1,\eta_2))]  \nonumber \\
                                         && ~~~ + G_2({\rm min}(\eta_1,\eta_2))[1-G_1({\rm max}(\eta_1,\eta_2))]  \nonumber \\
&&  ~~~ - \left ( G_1(\eta_1)[1-G_2(\eta_1)] + G_2(\eta_1)[1-G_1(\eta_1)] \right ) \nonumber \\
&& ~~~ \times \left ( G_1(\eta_2)[1-G_2(\eta_2)] + G_2(\eta_2)[1-G_1(\eta_2)] \right ) \big \}.    \label{eq:cov2} 
\end{eqnarray}
A simplification occurs along the diagonal $\eta_1=\eta_2=\eta$, where
\begin{eqnarray}
{\rm cov}(\eta,\eta) &=& \omega^2 \big \{ \tfrac{1}{4}\!-\! 4 [G_1(\eta)-\tfrac{1}{2}]^2 [G_2(\eta)-\tfrac{1}{2}]^2 \big \} \nonumber \\
&=& f(\eta)[\omega-f(\eta)].    \label{eq:diag} 
\end{eqnarray}
Also, the leading expansion at the diagonal in the anti-diagonal direction, with $\eta_1=\eta +\delta$ and  $\eta_2=\eta -\delta$, yields a very 
simple formula, 
\begin{eqnarray}
&& {\rm cov}(\eta\!+\!\delta,\eta\!-\!\delta)={\rm cov}(\eta,\eta) - \omega^2 [g_1(\eta)\!+\!g_2(\eta)] |\delta| + {\cal O}(\delta^2). \nonumber \\ \label{eq:delta}
\end{eqnarray}

Figure~\ref{fig:cov} shows the resulting distributions for ${\rm cov}(\eta_1,\eta_2)$ for the three considered cases.
One observes vivid qualitative differences between the covariances in cases i) and ii), cf. Figs.~\ref{fig:cov}a) and b). Whereas in case i) 
the covariance exhibits a monotonously increasing ridge along the $\eta_1=\eta_2$ direction, the covariance in case ii) shows a double peak structure, 
with a zero at $\eta=\eta_{\rm max} \simeq 2.5$, which corresponds to the zero of $g_1$ and $g_2$ in Fig.~\ref{fig:g1g2}a). At this point 
$G_1(\eta)=0$ and $G_2(\eta)=1$, which upon substitution to Eq.~(\ref{eq:cov2}) yields zero.
Another difference is in magnitude of the covariance, which in case i) is significantly larger than in case ii).

The covariance in case iii) is very close to case i) (cf. Figs.~\ref{fig:cov}a) and c)). 
Some small difference can be seen  where $\eta_1$ is small(large), but $\eta_2$ large(small),
where in case iii) the covariance noticeably drops to negative values.

We also note that in all cases the values on the diagonal is obeying Eq.~(\ref{eq:diag}). The fall-off from the diagonal in the anti-diagonal 
direction is given by the second term in Eq.~(\ref{eq:delta}). We note that the slope is proportional to \mbox{$4 f(\eta_{1,2}^{(0)})[g_1(\eta)+ g_2(\eta)]$}, 
hence two models which have similar values of  $\eta_{1,2}^{(0)}$ and close sums of the two end-point distributions, $g_1(\eta)+ g_2(\eta)$, will have 
similar covariances in the vicinity of the diagonal. Both conditions are satisfied between models i) and iii). In particular, we can see that the sum 
$g_1(\eta)+ g_2(\eta)$ for model iii) in Fig.~\ref{fig:g1g2}a) (dotted lines) is close to twice  $g_{1,2}(\eta)$ for model i) (solid line).

Thus the reason for the similarity of correlations in
cases (i) and (iii) may be traced back to Eq.~(\ref{eq:delta}),
which shows that this is the average of $g_1(\eta)$  and $g_2(\eta)$, which controls the fall-off of the
correlation from the diagonal. These averages happen to be very similar when we use any reasonable parametrization of
the parton distribution function giving the PDF of one end-point distribution, and the fluctuations the other end-point are adjusted 
to match the profile function $f(\eta)$, as explained in Sec.~\ref{sec:end}.

\section{Correlations from multiple strings \label{sec:corrm}}

As already discussed in the introduction, in our approach the strings ``belong'' to the valence 
quarks either from nucleus A or from nucleus B. 
With the underlying assumptions of independent wounded sources, the expressions for the 
$n$-body distributions account for the combinatorics in a simple manner, with the particles at rapidities $\eta_i$ 
being products from a string belonging to $A$ or to $B$.  For the one-body density 
in A-B collisions one finds
\begin{eqnarray}
{f_{AB}(\eta)} = \br{N_A}  {f_A(\eta)} + \br{N_B}  {f_B(\eta)}, \label{eq:onebody}
\end{eqnarray}
where  $\br{N_A}$ and $\br{N_B}$ are the event-by-event 
average numbers of wounded sources in nuclei A and B, respectively, 
and $f_{A,B}(\eta)=f(\pm \eta)$ denote the profiles for the emission from a single string, as given by Eq.~(\ref{eq:1}), associated with sources from nuclei $A$ or $B$. 
We work in the nucleon-nucleon center-of-mass (CM) frame, hence $f_{A}(\eta)=f_{B}(-\eta)$.

Analogously, one can define the two-body distribution for emission from a single string in nuclei $A$ and $B$ as $f_{A,B}(\eta_1,\eta_2)=f_2(\pm\eta_1,\pm\eta_2)$, 
and the corresponding covariances as ${\rm cov}_{A,B}(\eta_1,\eta_2)={\rm cov}(\pm\eta_1,\pm\eta_2)$.
Then, one readily obtains the covariance for the production in A-B collisions (see Appendix \ref{app:comb}) in the form
\begin{eqnarray}
&& {\rm cov}_{AB}(\eta_1, \eta_2)  \equiv f_{AB}(\eta_1,\eta_2) -  f_{AB}(\eta_1)  f_{AB}(\eta_2)   \nonumber \\
                            &=& \br{N_A} {\rm cov}_A(\eta_1,\eta_2) + \br{N_B} {\rm cov}_B(\eta_1,\eta_2)   \label{eq:gen} \\
                            &+& {\rm var}(N_A) {f_A(\eta_1)}{f_A(\eta_2)} + {\rm var}(N_B) {f_B(\eta_1)}{f_B(\eta_2)} \nonumber \\
                            &+& {\rm cov}(N_A,N_B) \left [{f_A(\eta_1)} {f_B(\eta_2)}+ {f_B(\eta_1)} {f_A(\eta_2)} \right ]. \nonumber
\end{eqnarray}
In the special case of symmetric collisions Eq.~(\ref{eq:gen}) simplifies into
\begin{eqnarray}
&& {\rm cov}_{AB}(\eta_1, \eta_2)  =  \br{N_A} {\rm cov}_A(\eta_1,\eta_2) + \br{N_B} {\rm cov}_B(\eta_1,\eta_2) \nonumber \\ 
&& +  {\rm var}(N_+) {f_s(\eta_1)} {f_s(\eta_2)} + {\rm var}(N_-) {f_a(\eta_1)} {f_a(\eta_2)},  \label{eq:gensym}
\end{eqnarray}
where $N_-=N_A-N_B$.
The moments of  $N_A$ and $N_B$ evaluated with {\tt GLISSANDO} are listed in Appendix~\ref{app:GL}.

\begin{figure}
\begin{center}
\includegraphics[width=0.48\textwidth]{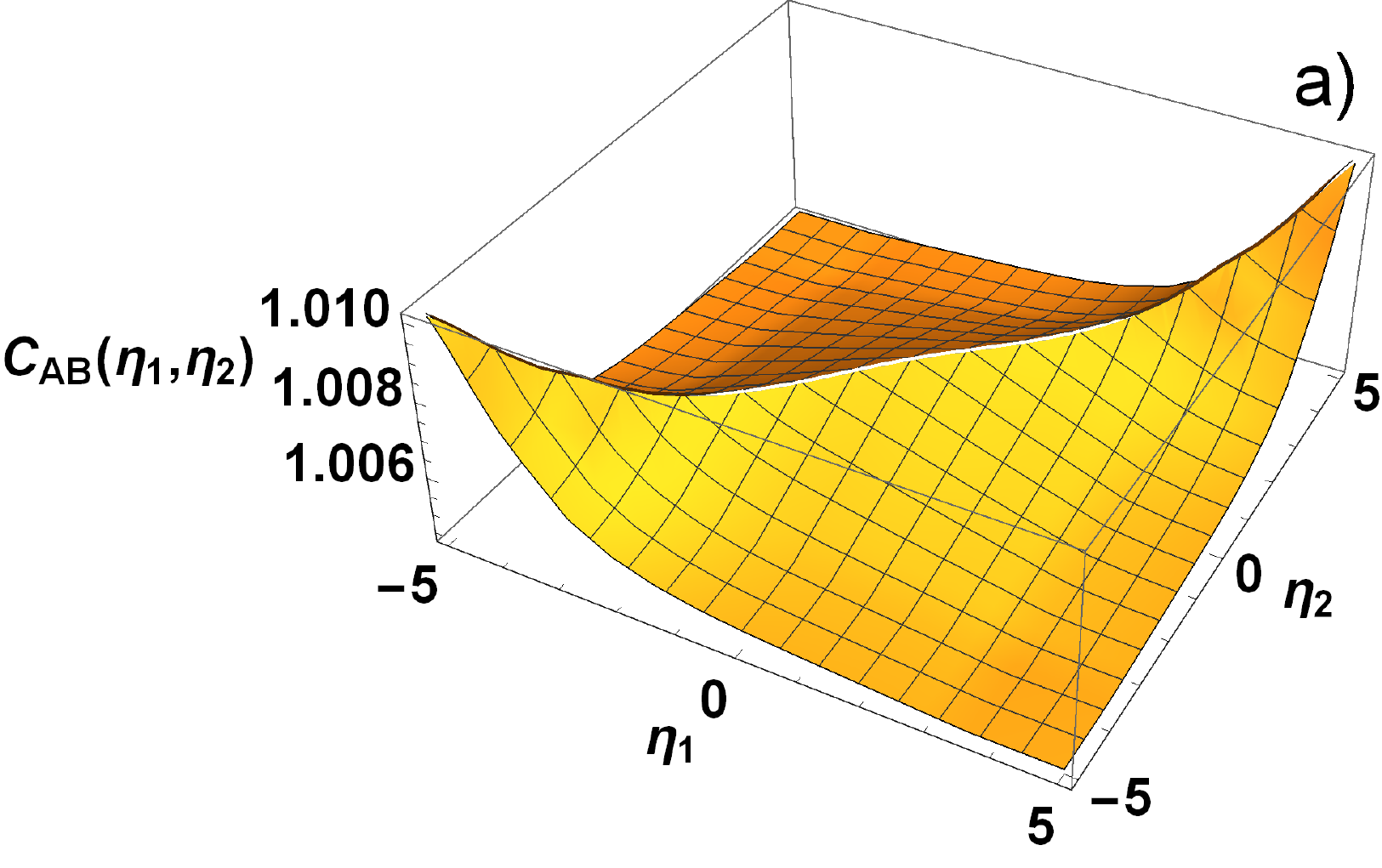}
\includegraphics[width=0.48\textwidth]{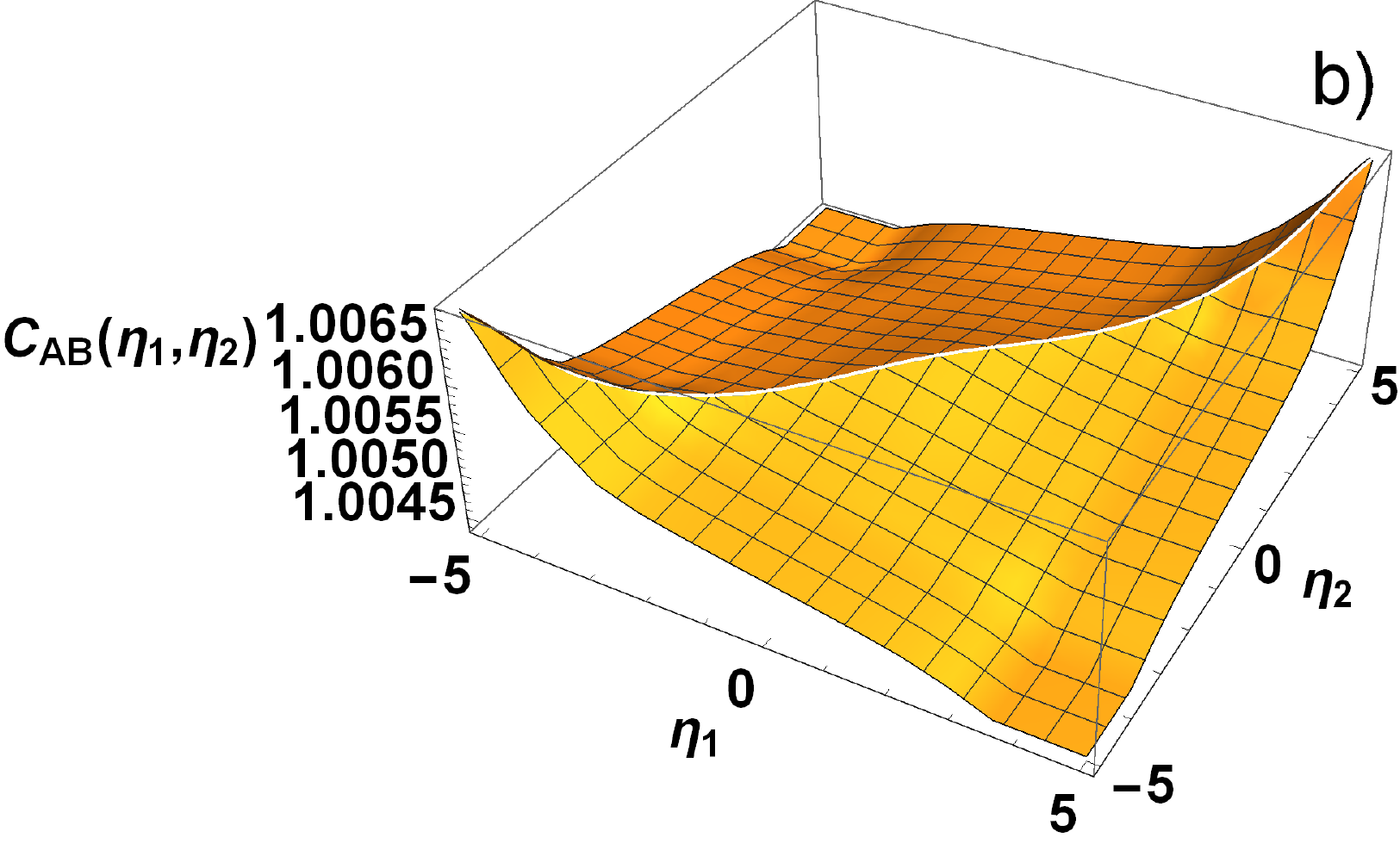}
\includegraphics[width=0.48\textwidth]{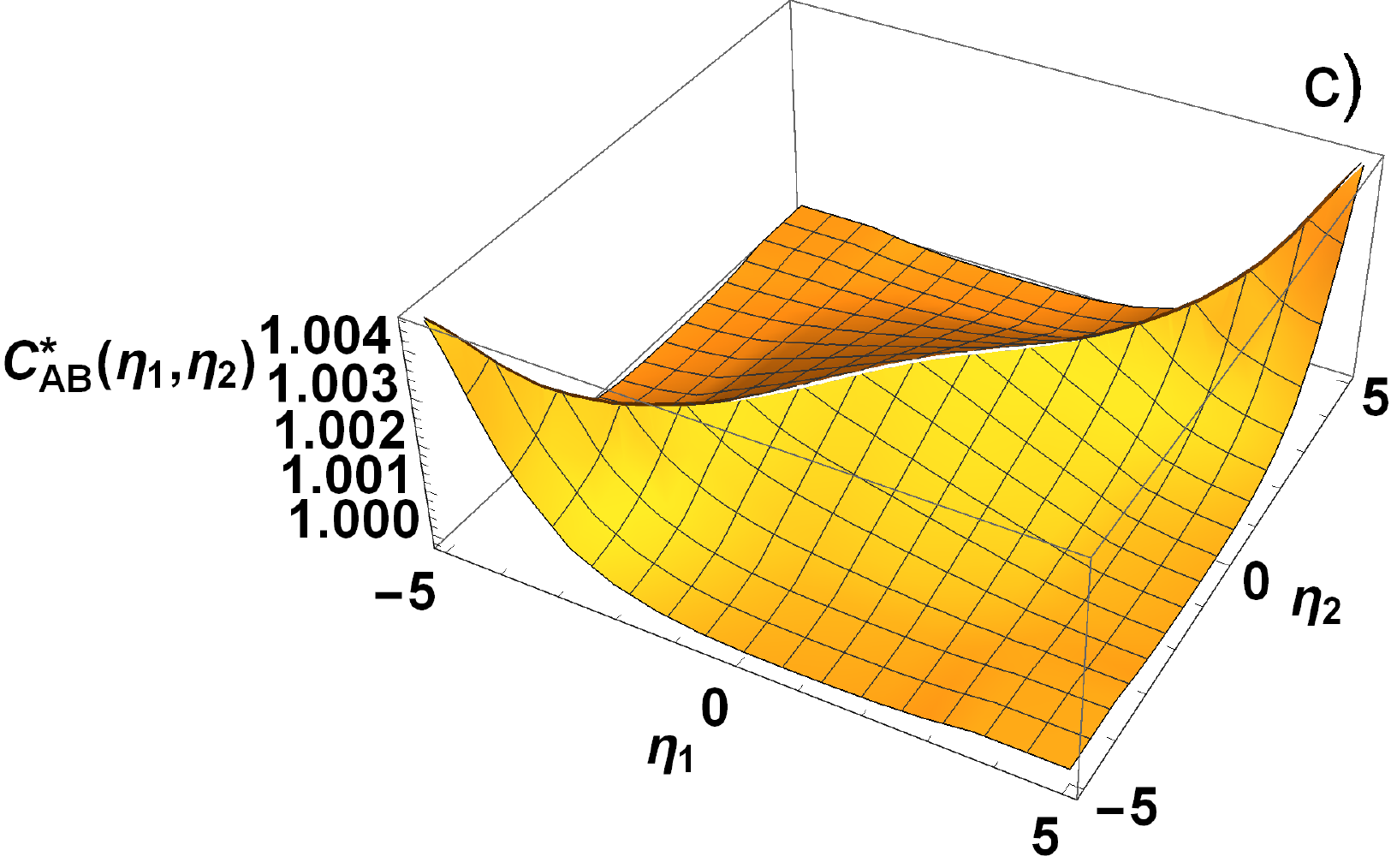}
\end{center}
\vspace{-5mm}
\caption{
Correlations $C_{AB}(\eta_1,\eta_2)$ for the $6\%$ most central Au-Au collisions for model cases i) (a) and  ii) (b), as well as $C^\ast_{AB}(\eta_1,\eta_2)$ 
for case i) (c). \label{fig:FBcorr}}
\end{figure}

We also introduce the customary correlation $C$ defined as 
\begin{eqnarray}
C_{AB}(\eta_1,\eta_2) =1+ \frac{{\rm cov}_{AB}(\eta_1,\eta_2)}{f_{AB}(\eta_1)f_{AB}(\eta_2)}, \label{eq:C}
\end{eqnarray}
which is a convenient measure due to its intensive property. 
For symmetric collisions Eq.~(\ref{eq:C}) becomes
\begin{eqnarray}
C_{AB}(\eta_1,\eta_2) =1+ \frac{{\rm cov}_{AB}(\eta_1,\eta_2)}{\br{N_+}^2 f_s(\eta_1)f_s(\eta_2)}, \label{eq:Csym}
\end{eqnarray}
To separate the contribution from the string end-point fluctuations, we also define
\begin{eqnarray}
C^\ast_{AB}(\eta_1,\eta_2) =  \frac{\br{N_A} {\rm cov}_A(\eta_1,\eta_2) + \br{N_B} {\rm cov}_B(\eta_1,\eta_2)  }{f_{AB}(\eta_1)f_{AB}(\eta_2)}. \nonumber \\
\label{eq:Cast}
\end{eqnarray}

We note that Eq.~(\ref{eq:gen}) or (\ref{eq:gensym}) contain terms with two classes of fluctuations: those stemming from single string end-point fluctuations, containing
${\rm cov}_i(\eta_1,\eta_2)$, which were the object of study in the previous section, 
and the remaining terms~\cite{Bzdak:2012tp} with moments of fluctuations of the numbers of wounded quarks, $N_A$ and $N_B$. 
Therefore 
the correlation function $C(\eta_1,\eta_2)$ contains a mixture of both effects. In principle, one could 
separate these effects via the technique of partial covariance  (see, e.g.,~\cite{Cramer:1946,krzanowski:2000}), 
which effectively imposes constraints on a multivariate sample. The details of such an analysis, which leads to very simple and practical 
expressions, were presented in~\cite{Olszewski:2017vyg}. 

In the present case, however, such an analysis is not necessary if we have in mind the standard $a_{nm}$ coefficients discussed in Sec.~\ref{sec:anm}.
As is clear from Eq.~(\ref{eq:Csym}), the term ${\rm var}(N_+) {f_s(\eta_1)} {f_s(\eta_2)}$ in  Eq.~(\ref{eq:gensym}) 
brings in a constant ${\rm var}(N_+)/\br{N_+}^2$ into $C(\eta_1,\eta_2)$. 
Therefore it only changes its baseline and does not affect the $a_{nm}$ coefficients (for $n,m\ge 0$). 
As we shall shortly see, the string end-point 
fluctuations given by the term with $\br{N_A} {\rm cov}_A(\eta_1,\eta_2) + \br{N_B} {\rm cov}_B(\eta_1,\eta_2)$
are largely dominant over the Bzdak-Teaney~\cite{Bzdak:2012tp} term, ${\rm var}(N_-) {f_a(\eta_1)} {f_a(\eta_2)}$, 
with the later entering at a level of 10-20\% in $a_{11}$ (cf. Sec.~\ref{sec:anm}). Hence  
one may simply take the view that measuring the $a_{nm}$ coefficients associated with $C(\eta_1,\eta_2)$ essentially provides information on 
the string end-point fluctuations, with 
only a small contamination by the fluctuation of the number of sources.

Panels a) and b) of Fig.~\ref{fig:FBcorr} show our results for $C_{AB}(\eta_1,\eta_2)$ of the $6\%$ most central Au-Au collisions 
at $\sqrt{s_{NN}}=200$~GeV in cases i) and ii) of our model. 
The correlations exhibit a ridge structure along the $\eta_1=\eta_2$ direction, which 
simply reflects the presence of the ridges in the single-string fluctuations displayed in Fig.~\ref{fig:cov}.
The correlation in case iii) is very close to case i), simply reflecting the behavior of Fig.~\ref{fig:cov}, hence we do not include it in the plot.

Panel c) shows the correlation stemming from the fluctuation of the string end-point, $C^\ast_{AB}(\eta_1,\eta_2)$ of Eq.~(\ref{eq:Cast}). 
We note that, apart for an overall shift by a constant, it
is very similar to the correlation  $C_{AB}(\eta_1,\eta_2)$ of Eq.~(\ref{eq:C}),
which indicates an important feature shown by our study:
 The shape of the correlation function $C(\eta_1,\eta_2)$ is largely dominated by the string end-point fluctuations, 
whereas the effects of the fluctuations of the number of sources are small.

\section{$a_{nm}$ coefficients \label{sec:anm}}

\begin{figure}
\hfill \includegraphics[width=0.48\textwidth]{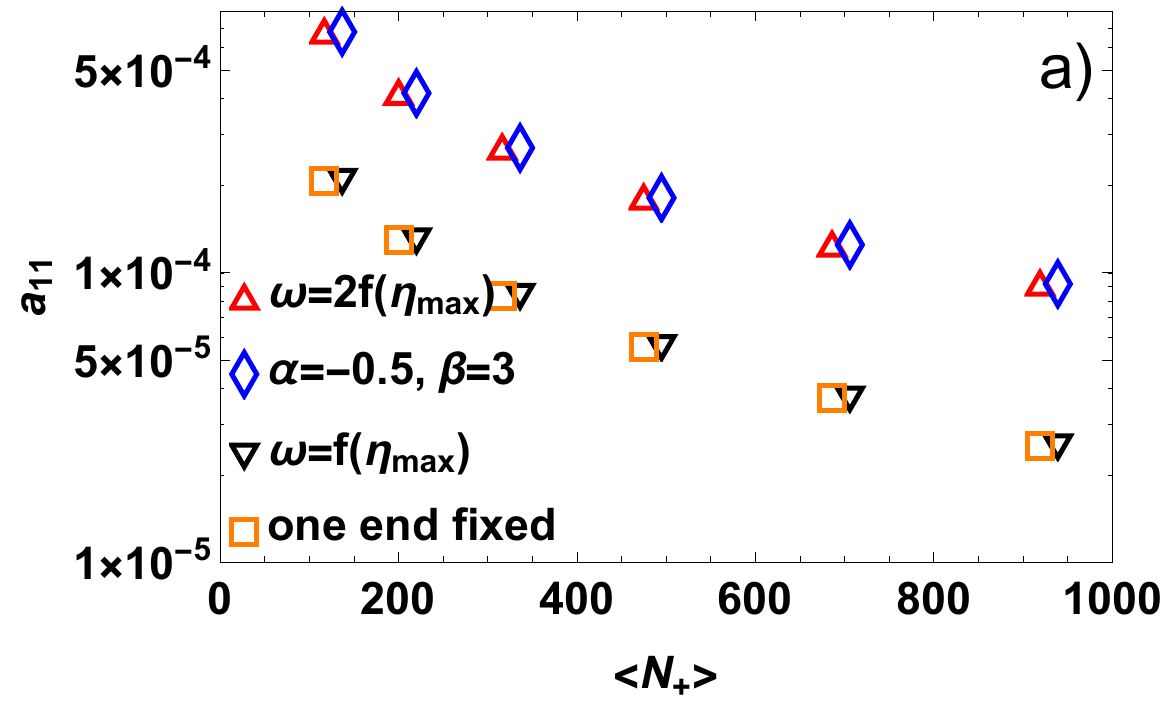}\\ \vspace{3mm}
\hfill \includegraphics[width=0.48\textwidth]{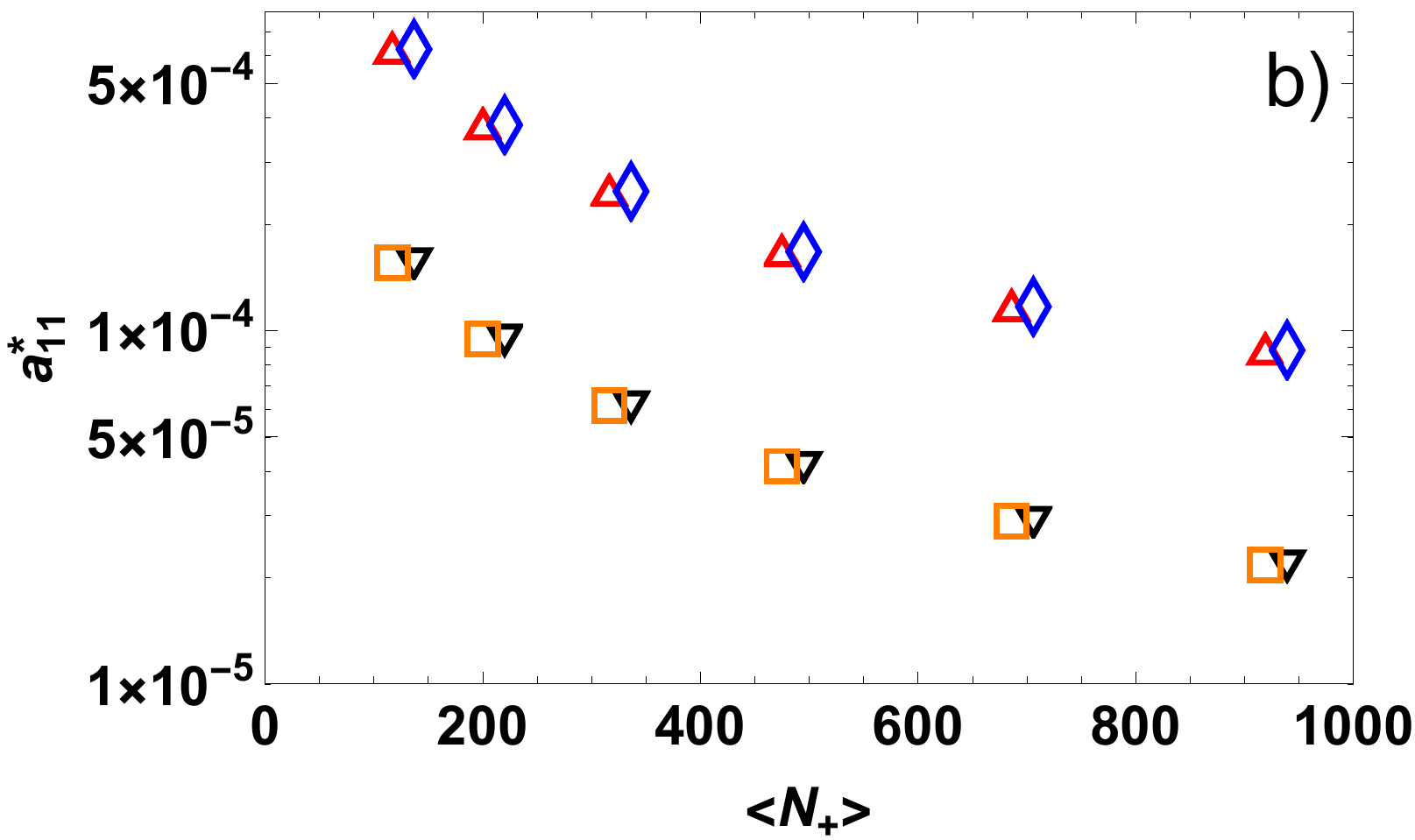}\\ \vspace{3mm}
\hfill \hspace*{-0pt}\includegraphics[trim={3.pt 0 0pt 0},clip,width=0.44\textwidth]{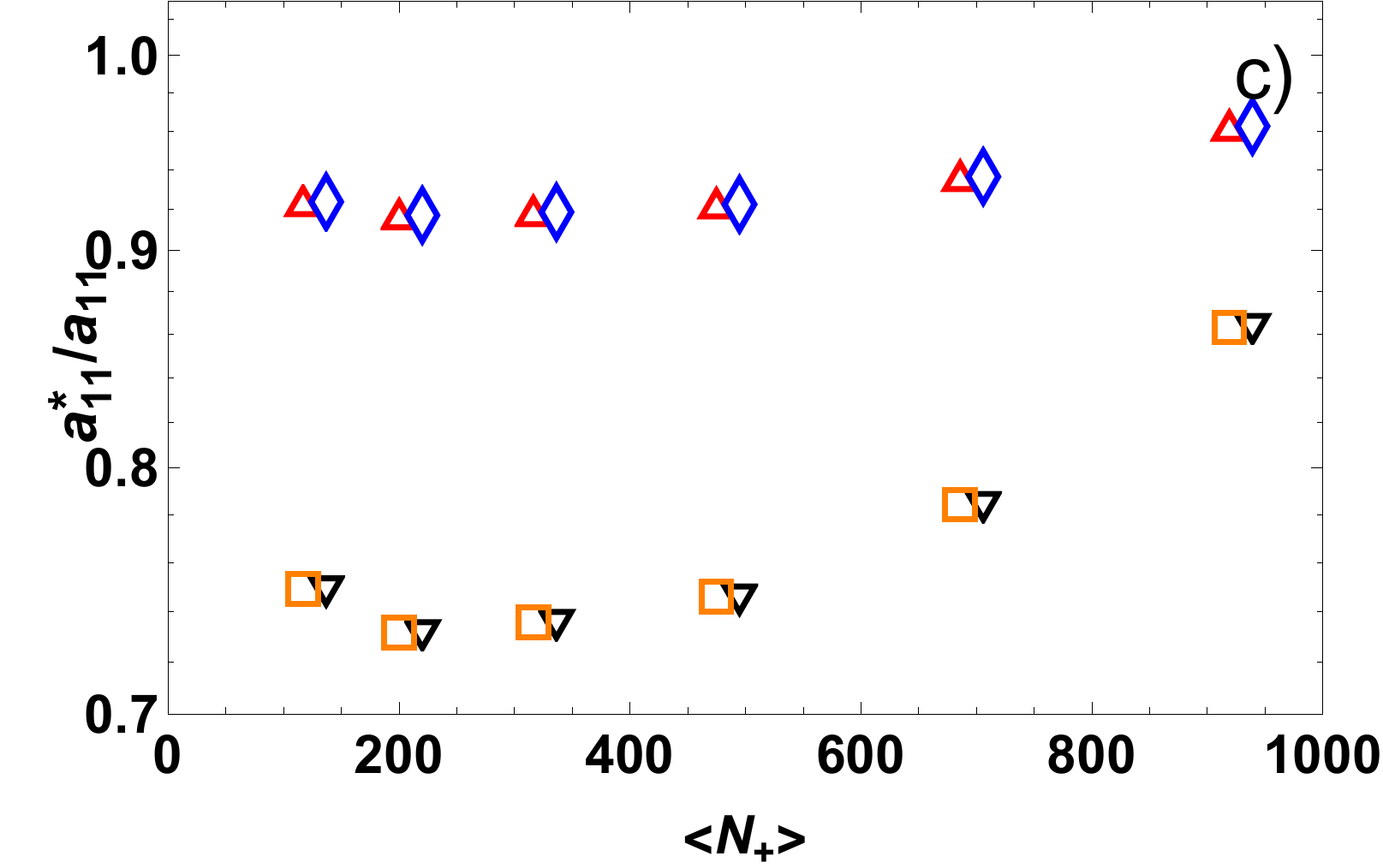}\:\!\:\!\\
\vspace{-2mm}
\caption{$a_{11}$ (a) and $a^\ast_{11}$ (b) for Au-Au collisions at $\sqrt{s_{\rm NN}}=200$~GeV as 
a function of $\br{N_+}$ (the selected values for $\br{N_+}$ correspond to the $6$ centrality classes $0-6\%$, $6-15\%$, $15-25\%$, $25-35\%$, 
$35-45\%$, and $45-55\%$) in cases i) with $\omega=2f(\eta_{\rm max})$, ii) with $\omega=f(\eta_{\rm max})$, iii) with $\alpha=-0.5,\,\beta=3$, together with the model of~\cite{Broniowski:2015oif} with one end point fixed, as indicated in the legend. Panel c) displays the ratio $a^\ast_{11}/a_{11}$.
To enhance visibility, the markers for the overlapping cases are slightly shifted to the left or right along the abscissa. \label{fig:a11}}
\end{figure}

For a given correlation function $C(\eta_1,\eta_2)$, the $a_{nm}$  coefficients are defined 
as~\cite{Bzdak:2012tp,ATLAS:2015kla,ATLAS:anm}
\begin{eqnarray}
a_{nm} &=& \int_{-Y}^Y \frac{d \eta_1}{Y} \int_{-Y}^Y \frac{d \eta_2}{Y}\frac{1}{\mathcal{N}_C} C(\eta_1,\eta_2)
T_n\left(\frac{\eta_1}{Y}\right) T_m\left(\frac{\eta_2}{Y}\right), \nonumber \\ \label{eq:anmC}
\end{eqnarray}
with the normalization constant
\begin{equation}
\mathcal{N}_C=\int_{-Y}^Y \frac{d \eta_1}{Y} \int_{-Y}^Y \frac{d \eta_2}{Y} C(\eta_1,\eta_2),
\end{equation}
where $[-Y,Y]$ is the covered pseudorapidity range. Having in mind the typical pseudorapidity acceptance at RHIC, we use $Y=1$. 
The functions $T_n(x)$ form a set of orthonormal polynomials.  
The choice used in~\cite{ATLAS:2015kla,ATLAS:anm,Jia:2015jga} is 
\begin{eqnarray}
T_n(x)=\sqrt{n+1/2}P_n(x),
\end{eqnarray}
where $P_n(x)$ are the Legendre polynomials.

Analogously, we define 
\begin{eqnarray}
a_{nm}^\ast &=& \int_{-Y}^Y \frac{d \eta_1}{Y} \int_{-Y}^Y \frac{d \eta_2}{Y}\frac{1}{\mathcal{N}_C} C^\ast(\eta_1,\eta_2)
T_n\left(\frac{\eta_1}{Y}\right) T_m\left(\frac{\eta_2}{Y}\right), \nonumber \\ \label{eq:anmCast}
\end{eqnarray}
which focuses on the fluctuations of the strings
(note that the normalization constant  ${\mathcal{N}_C}$ is evaluated with $C(\eta_1,\eta_2)$ as in Eq.~(\ref{eq:anmC})).

\begin{figure}
\begin{center}
\includegraphics[width=0.48\textwidth]{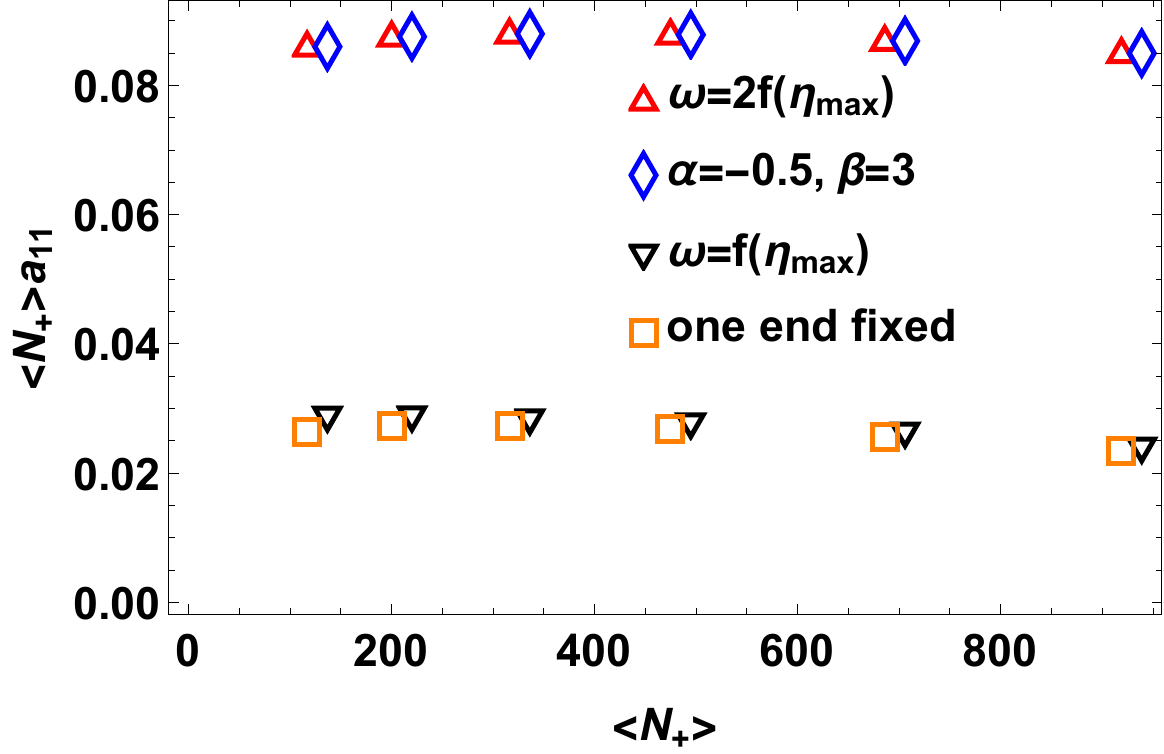}\\
\end{center}
\vspace{-5mm}
\caption{The product of $a_{11}$ and $\br{N_+}$, showing the scaling discussed in the text.  \label{fig:npa11}}
\end{figure}

Figure~\ref{fig:a11} shows our results for  $a_{11}$ (panel a) and $a^\ast_{11}$ (panel b) obtained for Au-Au collisions 
at $\sqrt{s_{\rm NN}}=200$~GeV and plotted as functions of the average 
number of wounded quarks $\br{N_+}$ in selected centrality classes. 
We note that the results for model cases i) and iii) are essentially identical, reflecting the feature seen already in Fig.~(\ref{fig:FBcorr}).
The result for case ii) is about a factor of 3 smaller. In this and following figures we also indicate the results for the model with single end-point fluctuations, 
which is identical to case ii) in the considered acceptance region.

In view of the discussion of Sec.~\ref{sec:end}, cases i) and ii) in Fig.~\ref{fig:a11} represent the upper and lower bounds for
the admissible values of the
$a_{11}$ coefficients. This is an important result, as it provides the possible range for this quantity in approaches sharing the 
features of our model. 

In panel c) of Fig.~\ref{fig:a11} we present the ratio $a^\ast_{11}/a_{11}$, which shows the announced dominance of the string end-point fluctuations 
over the fluctuation of the numbers of sources. In model cases i) and iii) the former account for 90\% of the effects, whereas in case ii) they account for 75-85\%.

From Eqs.~(\ref{eq:gensym},\ref{eq:Cast}) it is clear that $a^\ast_{11}$ scales  as $1/\br{N_+}$. 
For  $a_{11}$ there is a small departure of a relative order ${\rm var}(N_-)/\br{N_+}$. 
Numerically, for models i) and ii) $a^\ast_{11}\sim 0.08/\br{N_+}$, whereas the leading term of expansion (\ref{eq:delta}) yields a close 
result $a^\ast_{11}\sim 0.1/\br{N_+}$.
The approximate scaling for $a_{11}$ is exhibited in Fig.~\ref{fig:npa11}.

\begin{figure}
\begin{center}
\includegraphics[width=0.48\textwidth]{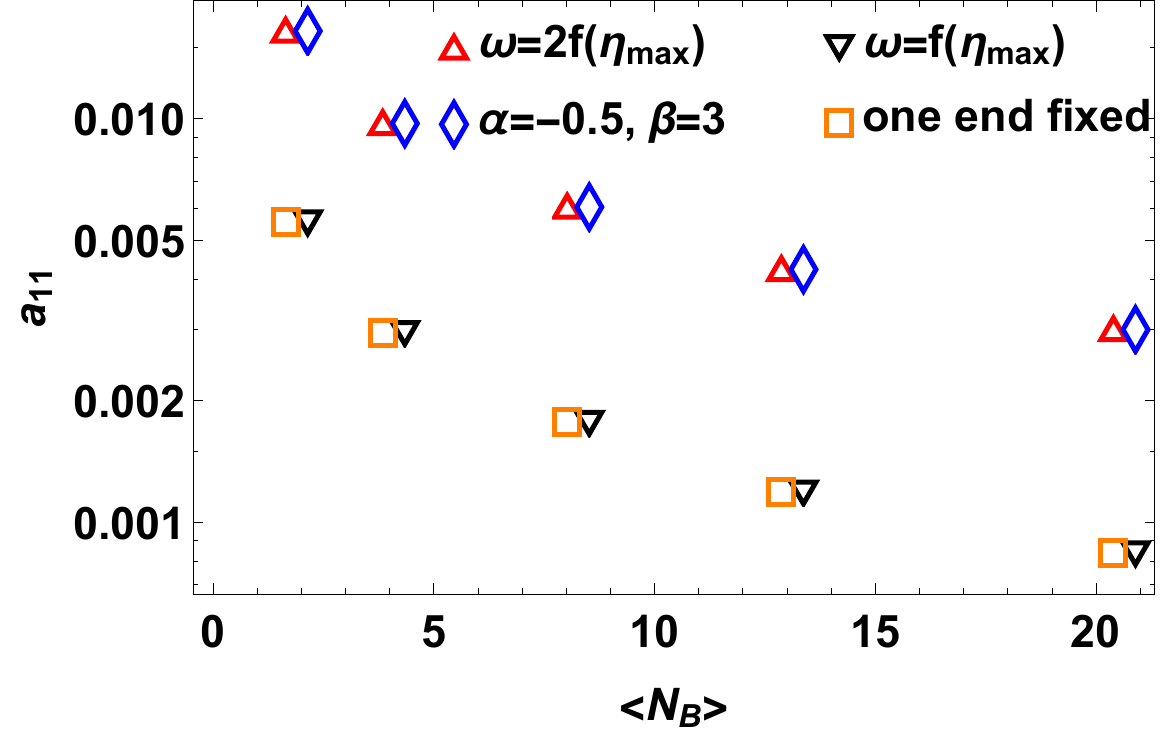}
\end{center}
\vspace{-5mm}
\caption{Same as in Fig.~\ref{fig:a11}a) but for d-Au collisions at $\sqrt{s_{\rm NN}}=200$~GeV, plotted as 
a function of the average number of wounded quarks in Au, $\br{N_B}$ (selected values for $\br{N_B}$ correspond to  
centrality classes $0-20\%$, $20-40\%$, $40-60\%$, $60-80\%$). \label{fig:a11dAu}}
\end{figure}

A similar analysis of the $a_{11}$ coefficients for the d-Au collisions yields qualitatively similar 
results, shown in Fig.~\ref{fig:a11dAu}. 
Here, the coefficients $a_{11}^\ast$ account for more than $90\%$ of the total, 
hence the dominance of string end-point fluctuations is even more pronounced in d-Au than in Au-Au collisions. 
For that reason we present only the results for $a_{11}$. 

\begin{figure}
\begin{center}
\includegraphics[width=0.48\textwidth]{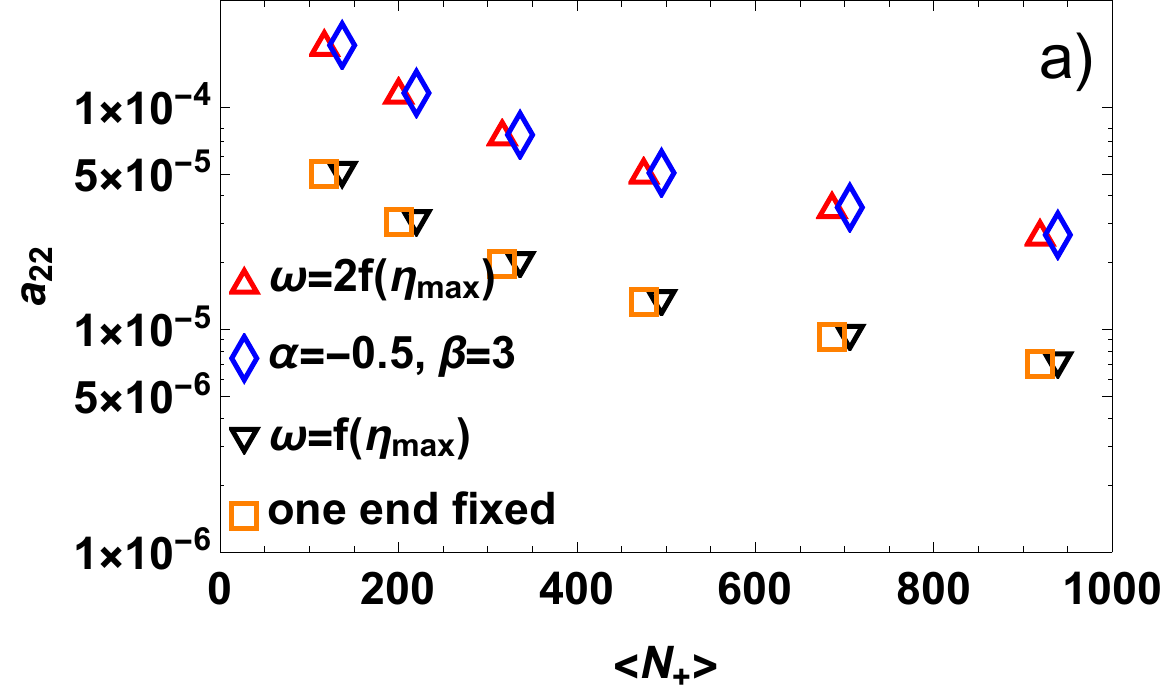}\\
\includegraphics[width=0.48\textwidth]{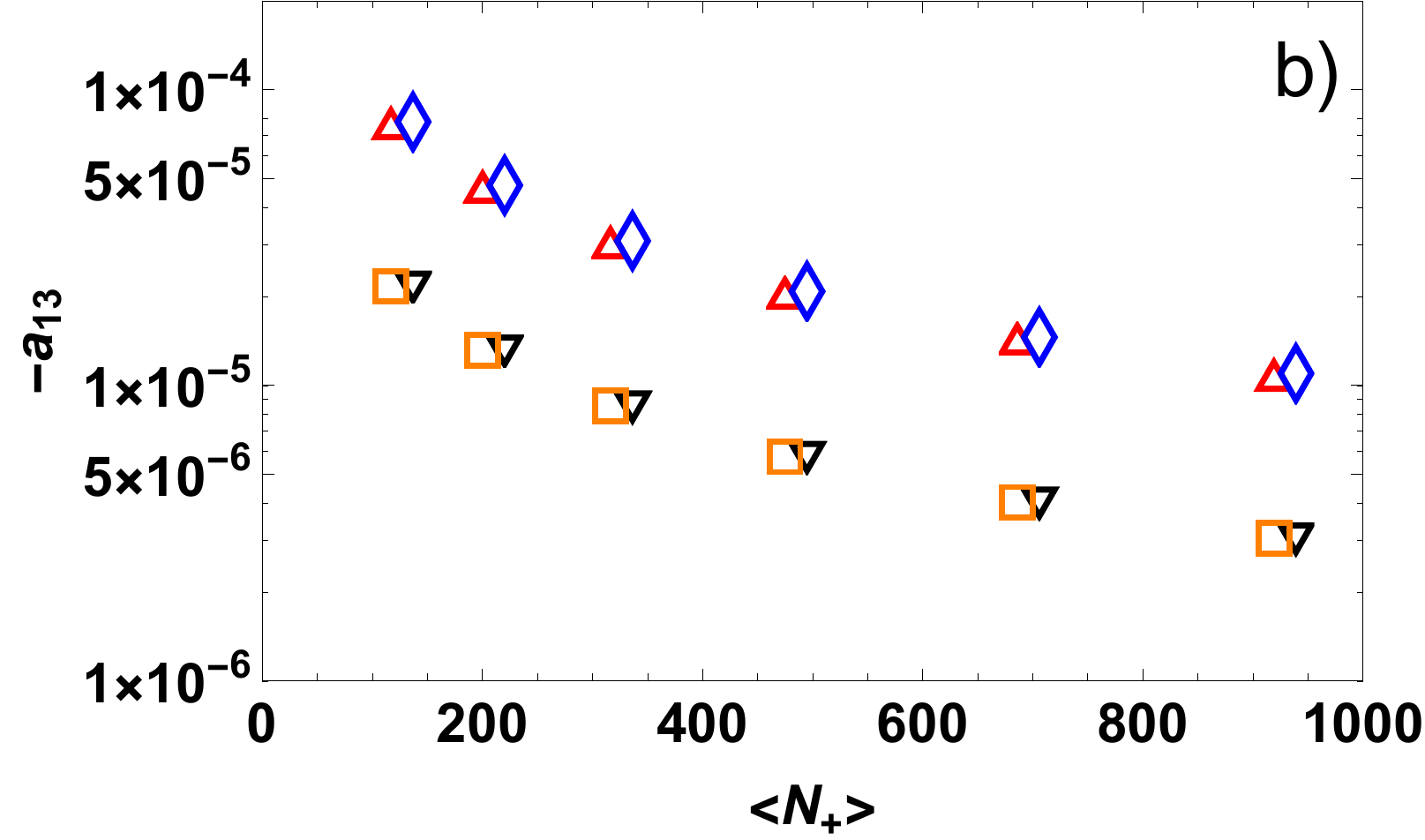}
\end{center}
\vspace{-5mm}
\caption{Same as in Fig.~\ref{fig:a11}a) but for $a_{22}$ (a) and $-a_{13}$ (b). \label{fig:anm}}
\end{figure}

In addition to $a_{11}$ coefficients, one may study the higher-order $a_{nm}$ coefficients. We 
give our results for $a_{13}$ and $a_{22}$ from Au-Au collisions 
in Fig.~ \ref{fig:anm}. While these coefficients are considerably suppressed as compared to $a_{11}$, shown in Fig.~\ref{fig:anm}, they exhibit the same qualitative behavior.
In particular, they scale almost exactly as $1/\br{N_+}$.

Finally, we remark that when the model results are to be compared to experimental values, one needs to relate the 
space-time rapidity of the initial stage, $\eta_{PS}=\frac{1}{2} \ln[(t+z)/(t-z)]$ (until now
denoted as $\eta$ in our considerations), to the 
momentum pseudorapidity of the measured hadrons, $\eta=\frac{1}{2} \log[(E+p_z)/(E-p_z)]$.
The experience of hydrodynamic simulations shows a mild longitudinal push, yielding $\eta \simeq 1.25 \,\eta_{PS}$. 
This effect leads to a quenching factor of about 1.5 to be applied to  the model $a_{nm}$ coefficients before comparing to the data.

\section{Conclusions \label{sec:concl}}

We have analyzed a model where strings are associated with wounded quarks and their end-points fluctuate.
We have used the data for the pseudo-rapidity spectra for d-Au and Au-Au collisions from the PHOBOS Collaboration at 
$\sqrt{s_{NN}}=200$~GeV to impose constraints on the one-body distributions in the model.
We have selected a RHIC energy for our study, since the wounded quark model works very well in this case. 

We first confirmed the results of~\cite{Barej:2017kcw} that a thus extracted one-body emission function reproduces reasonably well the  
experimental rapidity spectra and therefore  is universal in the sense that it can be applied to different centrality classes and collision systems 
for the considered collision energy.
Then we showed that there remains a substantial freedom in string end-point distributions $G_{1,2}$, which gives rise to a family of possible solutions.
Specifically, we have discussed three cases of solutions: the limiting cases i) and ii) and an intermediate case iii), inspired by the valence quark 
parton distribution function. We have argued that case ii) is equivalent to the model with single end-point fluctuations of~\cite{Broniowski:2015oif},
if the acceptance window at mid-rapidity is sufficiently narrow. 

The analysis was carried out analytically, which has its obvious merits.
We obtained formulas for the $n$-body distributions of the produced particles. 
In the study of the two-body correlations, we have examined the effects from string end-point fluctuations and from the 
fluctuation of the number of sources. The former largely dominate in the corresponding Legendre coefficients $a_{nm}$. 

We have found that the range for fluctuations is limited by two extreme cases. The lower limit, where the domains of the 
fluctuations of both ends do not overlap, coincides (for sufficiently narrow acceptance windows in pseudorapidity) 
with the model with single-end fluctuations considered earlier in~\cite{Broniowski:2015oif}. Allowing for both 
ends to fluctuate increases significantly the fluctuations, raising the $a_{nm}$ coefficients by a factor of $\sim 3$. 

A variant of the model where the distribution of one end of the string follows the valence quark PDF, 
is very close to the case giving maximum correlation (our case i)). 
Our results, in particular the presented bounds, 
can serve as a baseline for future data analysis of the forward-backward fluctuations in rapidity at $\sqrt{s_{NN}}=200$~GeV.

Our simple approach, while neglecting many possible effects such as mutual influence of the strings (merging into color ropes, nuclear shadowing), 
short range correlations of various origin, or 
assuming strings of only one type, incorporates two basic and generic features: fluctuation of the number of strings and fluctuation of the 
location of the string end-points. This makes its predictions valuable for understanding the underlying mechanisms.
It remains to be seen to what extent our analytic approach can be extended to more general models, 
in particular going beyond the simple Glauber wounded picture.

\begin{acknowledgments}
Research supported by the Polish National Science Centre (NCN) Grant No. 2015/19/B/ST2/00937.
\end{acknowledgments}

\appendix

\section{Matching the cumulative distribution functions to one-body emission profiles \label{app:match}}

It is convenient to introduce the shifted CDFs
\begin{eqnarray}
H_i(\eta)=G_i(\eta)-\frac{1}{2}, 
\end{eqnarray}
which grow from the value $-1/2$ up to $1/2$. 
Then Eq.~(\ref{eq:f1G}) can be rewritten as
\begin{eqnarray}
H_1(\eta) H_2( \eta)=\frac{1}{4}- \frac{1}{2\omega} f(\eta). \label{eq:h}
\end{eqnarray}

We shall now consider three specific cases.\footnote{We assume in the derivation of the first two cases that $f(\eta)$ is unimodal, as is the case 
of the phenomenologically fitted profile.}  In the first case, the maximum of $f(\eta)$ is taken to be $\omega/2$, which is the lowest possible 
value (otherwise it would contradict Eq.~(\ref{eq:1o2})). The position of the maximum is at $\eta_0 = \eta^{(1)}_0 = \eta^{(2)}_0$ (the two zeros of $H_i(\eta)$
coincide in this case). Then the solution takes the form 
\begin{eqnarray}
H_1(\eta)&=&\sqrt{\frac{1}{4}- \frac{1}{2\omega} f(\eta)}\,{\rm sgn}(\eta-\eta_0) s(\eta), \nonumber \\
H_2(\eta)&=&\sqrt{\frac{1}{4}- \frac{1}{2\omega} f(\eta)}\,{\rm sgn}(\eta-\eta_0) /s(\eta), \label{eq:case1}
\end{eqnarray}
where ${\rm sgn}$ denotes the sign function, and $s(\eta)$ is an arbitrary function chosen in such a way that the required 
limiting and monotonicity properties of $H_i(\eta)$ are preserved (one possibility, which we use, is $s(\eta)=1$, in which case both distributions are the same).

The second special case is when the maximum of $f(\eta)$ is $\omega$, which is the largest possible value. Then one may choose
\begin{eqnarray}
H_1(\eta)&=& - \frac{1}{2} \theta(\eta_0-\eta) +\left [ \frac{1}{2} - \frac{1}{\omega} f(\eta) \right ] \theta(\eta-\eta_0), \nonumber \\
H_2(\eta)&=& - \left [ \frac{1}{2} - \frac{1}{\omega} f(\eta) \right ] \theta(\eta_0-\eta)  + \frac{1}{2} \theta(\eta-\eta_0). \nonumber \\ \label{eq:case2}
\end{eqnarray}
In this case the supports of $g_1(\eta)$ and $g_2(\eta)$ are disjoint.

We can now easily verify that the formulas (\ref{eq:case1}) and (\ref{eq:case2}) indeed satisfy Eq.~(\ref{eq:h}).

In the intermediate case, when the maximum satisfies $\omega/2 < f(\eta) \le \omega$, one may generically take a ``favorite'' form of $H_1(\eta)$ and 
then evaluate $H_2(\eta)$ from Eq.~(\ref{eq:h}) as
\begin{eqnarray}
H_2(\eta)=\frac{\frac{1}{4}- \frac{1}{2\omega} f(\eta)}{H_1(\eta)}. \label{eq:h2}
\end{eqnarray}
Note that $H_2(\eta)$ is well-behaved near $\eta_1$, as in its vicinity
\begin{eqnarray}
&& H_1(\eta)=C_1^2 (\eta-\eta_1)+\dots, \nonumber \\
&& \frac{f(\eta)}{\omega}=\frac{1}{2}-C_2^2 (\eta-\eta_1)^2+\dots, 
\end{eqnarray}
where $C_1^2$ and $C_2^2$ denote positive constants,
hence 
\begin{eqnarray}
H_2(\eta)=\frac{C_2^2}{2C_1^2} (\eta-\eta_1)+\dots .
\end{eqnarray}
One needs to check explicitly if $H_2(\eta)$ obtained from Eq.~(\ref{eq:h2}) is a growing function, otherwise the initial choice of $H_1(\eta)$ is inconsistent.

Since $-\tfrac{1}{2} \le H_1(\eta) \le \tfrac{1}{2}$, it follows immediately from Eq.~(\ref{eq:h2}) that 
\begin{eqnarray}
H_2(\eta) \ge \frac{1}{2}- \frac{1}{\omega} f(\eta) {\rm ~~for~} \eta \ge \eta_0, \nonumber \\
H_2(\eta) \le -\frac{1}{2}-\frac{1}{\omega} f(\eta) {\rm ~~for~} \eta \le \eta_0 \label{eq:h3}
\end{eqnarray}
(and similarly for $H_1$), hence the expressions (\ref{eq:case2}) provide upper and lower limits for {\em any} CDF for the considered problem.

\section{PDF-motivated distribution \label{sec:pdf}}

When the string end-points $y_{1,2}$ are  associated with subnucleonic constituents, such as a valence or sea quark, gluon, or diquark, then 
they carry the fractions $x_{iA}$ or $x_{iB}$ of the  longitudinal momenta of the nucleons inside beams $A$ and $B$, respectively. 
Specifically, if the momentum of the constituent is 
$k_{iA}$ ($k_{iB}$) and the momentum of the nucleon is $P_A$ ($P_B$), then from standard kinematic 
considerations the corresponding rapidity $y_{iA}$ ($y_{iB}$) of the end-point is related 
to $x_{iA}$ ($x_{iB}$) with the exact formula
\begin{eqnarray}
&& x_{iA} \equiv \frac{k_{iA}^+}{P_A^+}=\frac{m_{Ti}}{M} e^{y_{iA}-y_b}, \nonumber \\
&& x_{iB} \equiv \frac{k_{iB}^+}{P_B^+}=\frac{m_{Ti}}{M} e^{-y_{iB}-y_b},
\end{eqnarray}
where $m_{T\,i}=\sqrt{m_i^2+k_{T\,i}^2}$ is the transverse mass of the constituent, $M$ is the mass of the nucleon, and $y_b$ is the rapidity of beam 
$A$ (in the assumed CM frame of the nucleon-nucleon collision, $-y_b$ is the rapidity of beam $B$).

The distributions of the locations of the string end-points are then defined via partonic distributions $p_i(x)$ as follows:
\begin{eqnarray}
g_i(y_{iQ}) dy_{iQ} = p_i(x_{iQ}(y_{iQ})) dx_{iQ},  \label{eq:PDFs}
\end{eqnarray}
with $Q=A,B$, or for the corresponding CDFs
\begin{eqnarray}
G_i(y_{iQ}) = P_i(x_{iQ}(y_{iQ})),. \label{eq:CDFs}
\end{eqnarray}

Since $x_{i\,Q}\in [0,1]$, the limits for the rapidities of the end points are $y_{iA}\in (-\infty,y_{i\uparrow}]$ and $y_{iB}\in [-y_{i\uparrow},\infty)$, where 
\begin{eqnarray}
y_{i\uparrow}=y_b+\log \left( \frac{M}{m_{Ti}}\right ).
\end{eqnarray}
In the CM reference frame of the nucleon-nucleon collision, the rapidity of the beam is
\begin{eqnarray}
 y_b =  \log \frac{\sqrt{s/4}+\sqrt{s/4-M^2}}{\sqrt{s/4}-\sqrt{s/4-M^2}} \simeq \log \frac{\sqrt{s}}{M},
\end{eqnarray}
therefore at $\sqrt{s} \gg M$ we have to a good approximation $y_{i\,\uparrow} \simeq \log \frac{\sqrt{s}}{m_{Ti}}$.

In the example used in this paper, a simple parametrization of the parton 
distribution functions (PDF) is used. Following many phenomenological studies, we take
\begin{equation}
p(x)=Ax^\alpha (1-x)^\beta,
\label{eq:pdffit}
\end{equation}
with the corresponding CDF
\begin{equation}
P(x)=\frac{B(x,1+\alpha,1+\beta)}{B(1,1+\alpha,1+\beta)},
\label{eq:B}
\end{equation}
where $B(z,a,b)$ denotes the incomplete Euler Beta function.

\begin{table}[tb]
\caption{First few moments of the wounded quark numbers in Au-Au collisions at $\sqrt{s_{NN}}=200$~GeV as 
obtained from {\tt GLISSANDO} simulations. The chosen centrality classes correspond to those in the PHOBOS experiment. \label{tab:AuAu}}
\vspace{2mm}
\begin{tabular}{c||r|r|r}
Centrality [\%]&$\langle{N_+}\rangle$&${ \rm var}(N_+)$& ${\rm var}(N_-)$ \\ \hline
0-6&929&4280&502\\
6-15&696&4649&653\\
15-25&484&2972&563\\
25-35&326&1472&399\\
35-45&210&811&262\\
45-55&126&396&144\\\hline
\end{tabular}
\end{table}

\begin{table}[tb]
\caption{Same as in Table~\ref{tab:AuAu} but for d-Au collisions at $\sqrt{s_{NN}}=200$~GeV. 
Here $N_A$ and $N_B$ denote the number of wounded quarks in d and Au, respectively. \label{tab:dAu}}
\vspace{2mm}
\begin{tabular}{c||r|r|r|r|r}
Centrality [\%]&$\langle{N_A}\rangle$&$\langle{N_B}\rangle$&${ \rm var}(N_A)$& ${\rm var}(N_B)$& ${\rm cov}(N_A,N_B)$\\\hline
0-20&5.9& 20.6& 0.1& 14.8& 0.1\\
20-40&5.3&13.1&0.8&2.8&-0.3\\
40-60&4.1& 8.3& 1.0& 2.7& -0.4\\
60-80&2.8& 4.1& 0.6& 1.4& 0.0\\
80-100&1.6& 1.9&0.3&0.3& -0.1\\\hline
\end{tabular}
\end{table}

\section{$2$-body density \label{app:comb}}

When we consider the two-body density of particles produced from multiple strings formed in A-B collisions, there are several 
combinatorial cases which may occur: the two particles may originate from the same string associated with A, 
from different strings associated with A,  from the same string associated with B, 
from different strings associated with B, and finally one particle is emitted from a string associated with A and the other from as string 
associated with B. Thus, the two-body density averaged over events in A-B collisions takes the form 
\begin{align}
&f_{AB}(\eta_1,\eta_2)=
 \nonumber\\&
\br{N_A}f_{A}(\eta_1,\eta_2)+\br{N_A(N_A-1)}f_{A}(\eta_1)f_{A}(\eta_2)
 \nonumber\\&
+\br{N_B}f_{B}(\eta_1,\eta_2)+\br{N_B(N_B-1)}f_{B}(\eta_1)f_{B}(\eta_2)
 \nonumber\\&
+\br{N_AN_B}(f_{A}(\eta_1)f_{B}(\eta_2)+f_{B}(\eta_1)f_{A}(\eta_2))
\,,
\end{align}
We define the covariances in the usual way,
\begin{align}
{\rm cov}_A(\eta_1,\eta_2)&=f_A(\eta_1,\eta_2)-f_A(\eta_1)f_A(\eta_2)\,,\nonumber\\
{\rm cov}_B(\eta_1,\eta_2)&=f_B(\eta_1,\eta_2)-f_B(\eta_1)f_B(\eta_2).
\end{align}
Then
\begin{align}
&f_{AB}(\eta_1,\eta_2)=
 \nonumber\\&
\br{N_A}{\rm cov}_{A}(\eta_1,\eta_2)+\br{N_A^2}{f_{A}(\eta_1)}{f_{A}(\eta_2)}
 \nonumber\\&+
\br{N_B}{\rm cov}_{B}(\eta_1,\eta_2)+\br{N_B^2}{f_{B}(\eta_1)}{f_{B}(\eta_2)}
 \nonumber\\&
+\br{N_AN_B}({f_{A}(\eta_1)}{f_{B}(\eta_2)}+{f_{B}(\eta_1)}{f_{A}(\eta_2)}),
\end{align}
and Eq.~(\ref{eq:gen}) follows.

\section{Moments of the wounded quark distributions \label{app:GL}}

The lowest moments of the wounded quark distributions obtained form {\tt GLISSANDO}~\cite{Broniowski:2007nz,Rybczynski:2013yba} 
and used in our analysis are 
collected in Tables~\ref{tab:AuAu} and \ref{tab:dAu}.

\FloatBarrier

\end{document}